\RequirePackage{lineno} 

\documentclass[twocolumn,preprintnumbers,amsmath,amssymb,nofootinbib]{revtex4}
\usepackage{graphicx}% Include figure files
\usepackage{dcolumn}% Align table columns on decimal point
\usepackage{bm}% bold math

\usepackage{epstopdf}
\def\bea{\begin{eqnarray}}
\def\eea{\end{eqnarray}}

\def\pp{\mbox{$p$-$p$}}

\def\pa{\mbox{$p$-A}}

\def\auau{\mbox{Au-Au}}

\def\pbpb{\mbox{Pb-Pb}}
\def\ppb{\mbox{$p$-Pb}}

\def\aa{\mbox{A-A}}

\def\ee{\mbox{$e^+$-$e^-$}}

\def\qqbar{\mbox{$q$-$\bar q$}}
\def\pt{$p_t$}
\def\pz{$p_z$}
\def\mt{$m_t$}
\def\yt{$y_t$}
\def\yz{$y_z$}
\def\mpt{$\langle p_t \rangle$}
\def\nch{$n_{ch}$}
\def\mmpt{$\bar p_t$}
\def\etaa{$\eta$}

\begin{document} 
	
	\setlength{\pdfpagewidth}{8.5in}
	\setlength{\pdfpageheight}{11in}
	
	\setpagewiselinenumbers
	\modulolinenumbers[5]
	
	\addtolength{\footnotesep}{-10mm}\
	% \linenumbers

\preprint{version 1.3\textsl{}}

\title{Comparison of spectrum models as applied to single-particle $\bf p_t$ spectra from high-energy p-p collisions and their physical interpretations
}

\author{Thomas A.\ Trainor}\affiliation{University of Washington, Seattle, WA 98195}

\date{\today}

\begin{abstract} 

A parametrized mathematical model is required to extract the information carried by transverse momentum $p_t$ spectra from high-energy nuclear collisions and subject it to physical interpretation in terms of possible hadron production mechanisms. It is essential that model elements correspond accurately to relevant mechanisms. The importance of proper model construction and implementation has increased with the emergence of claims for ``collectivity'' (flows) associated with small collision systems (e.g.\ \pp\ and \ppb). Argument by analogy concludes that if flows demonstrate quark-gluon plasma (QGP) formation in \aa\ collisions, as currently believed, then QGP formation must also occur in small systems despite low particle densities. A two-element spectrum model, denoted herein as the Bylinkin model, includes an exponential element and a ``power-law'' element interpreted by the authors to represent emission from a thermalized source and from jet production respectively. Application of the Bylinkin model to various collision systems has led to conclusions about achievement of thermalization and other characteristics of nuclear dynamics. In connection with the Bylinkin model there has emerged theoretical conjecture that the thermalization mechanism signaled by the exponential element is hard processes interacting with quantum entanglement within projectile protons. Predating the Bylinkin model is a two-component (soft+hard) model (TCM) derived empirically from the evolution of \pp\ spectrum data with event multiplicity as a form of data compression. The TCM has been applied to many collision systems and hadron species from which a systematic description of high-energy nuclear collisions has emerged. The Bylinkin model can be seen as a limiting case of the TCM model functions that is unsuited to represent underlying production mechanisms. The present study provides detailed comparisons of the two models for a variety of situations and contrasts two very different data interpretations that result.

\end{abstract}

\maketitle

%%%%%%%%%%%%%
\section{Introduction}

Inferred presence of collectivity (flows) within high-energy collisions involving small systems (e.g.\ \pp, \pa\ collisions) since commencement of LHC operations in 2010 has been interpreted to support claims of quark-gluon plasma (QGP) formation in small systems~\cite{nagle}. The claims are based  on certain data features in spectra and angular correlations including ``hardening'' of spectra with increasing event multiplicity \nch\ and hadron mass that appear similar to spectrum trends in \aa\ collisions where QGP formation is considered to be established~\cite{alicepppid}. Such claims are counterintuitive based on conventional understanding of QCD and thus merit careful examination of arguments and evidence in their favor.

Reference~\cite{alicepppid} observes that at low \pt\ ``...collective phenomena are observed in...[\pp, \ppb\ and \pbpb] collisions....'' It asserts that ``...in order to describe bulk particle production in \aa\ collisions, one usually relies on hydrodynamic and thermodynamic modeling....'' 
Spectrum variation that may be associated with radial expansion (flow) is ``....studied in the context of the Boltzmann-Gibbs Blast-Wave [BW] model.'' Parameter values derived from BW model fits to \pt\ spectra are  interpreted as indicating thermal temperatures and flow velocities.

If QGP formation in small collision systems is accepted based on interpretation of certain collision data then the problem emerges how a thermalized dense medium, assumed as a precursor to QGP formation, is achieved in small low-density systems. One possibility involves the role of {\em quantum entanglement} in high-energy collisions. A theoretical conjecture proposes that entanglement might be manifested in the entropy of the hadronic final state of deep-inelastic-scattering (DIS) data~\cite{dimalev}. The argument has been extended to propose quantum entanglement as the mechanism for thermalization in small collisions indicated by observation of certain spectrum features~\cite{dimaentangle}.

Just as BW model fits to \pt\ spectra have been interpreted to demonstrate the presence of radial flow and to provide estimates of thermal temperature and radial speed of a flowing medium, a spectrum model composed of two  functions has been interpreted to indicate thermalization based on one of its elements. The model consists of combined exponential and power-law elements, referred to here as the Bylinkin model after its principal author~\cite{bylinkin2010}. As the model is described the presence {\em or absence} of its exponential element, as required for description of collision data, determines whether a collision system has thermalized in the Boltzmann sense of multiple particle collisions that  tend to maximize entropy.

The Bylinkin model has been applied to a number of collision systems (e.g.~\cite{bylinkin2010,bylinkin2014nucphb,bylinkinanom}) with inference of parameters $T_{th}$ for the exponential element and $T_h$ and $N$ for the power-law element. Parameters $T_{th}$ and $T_h$ are observed to be linearly correlated, which coupling has been interpreted to support the hypothesis of thermalization via quantum entanglement~\cite{pajares}. The ratio of amplitudes for exponential and power-law elements exhibits markedly different behavior for different collision systems, and the exponential element appears to be absent for proton spectra from \pp\ and \auau\ collisions~\cite{bylinkinanom}.

Just as application of the BW model to spectra can be questioned~\cite{tommodeltests} the appropriateness and interpretability of the Bylinkin model should be closely examined, especially given the significance of inferences derived from it. In this study the Bylinkin model is compared to a two-component (soft+hard) model (TCM) applied to many collision systems with considerable success~\cite{ppprd,alicetommpt,ppquad,ppbpid}. In contrast  to the Bylinkin model the TCM is not based on {\em a priori} assumptions. It is instead derived empirically from evolution of data spectra over a wide range of collision systems and energies~\cite{alicetomspec} and can be viewed as a form of {\em lossless data compression}~\cite{tommodeltests}. In what follows the Bylinkin model is placed in the context of model development over several decades, model-model and model-data comparisons are presented, and the relation of \pt\ spectrum models to the question of thermalization is reviewed. As a general conclusion (a) the Bylinkin model is a particular case of the more-general TCM, (b) neither its exponential nor power-law elements is {\em required} by data and (c)  its parameters are not physically interpretable as they are intended. Given those results the scenario that quantum entanglement leads to thermalization or even the presence of thermalization within collisions itself, if based on that model, can be questioned.

%%%%%%%%%%%%%%%%
This article is arranged as follow
Section~\ref{tcmmod} briefly summarizes the history of \pt\ spectrum models applied to data from high-energy collisions, including combinations of model elements.
Section~\ref{compare} provides a detailed study of data-model and model-model comparisons.
Section~\ref{compress} describes derivation of a TCM for densities on \pt\ and $\eta$ empirically without {\em a priori} assumptions as a form of data compression.
Section~\ref{predict} contrasts the predictivity of models based on their ability to describe new data based on extrapolation from previous analysis.
Section~\ref{thermal} considers the relation of thermalization and entanglement to spectrum models.
Sections~\ref{disc} and~\ref{summ} present discussion and summary. 
%Appendix~\ref{app}

%%%%%%%%%%%%%
\section{Two-component spectrum models} \label{tcmmod}

This section provides a brief history of \pt\ spectrum model development for high-energy nuclear collisions, describes empirical derivation of a two-component model for \pp\ collisions based on evolution of spectrum shape with event multiplicity \nch\ and finally describes a two-element model based on certain {\em a priori} assumptions.

\subsection{$\bf p_t$ spectrum model development}

The structure of \pt\ spectra near midrapidity in high-energy hadron-hadron and hadron-nucleus collisions has long been a subject of interest. Soon after commencement of the National Accelerator Lab (NAL, now FNAL), fixed-target $p$-A experiments revealed unexpectedly high particle yields at high \pt, i.e.\ spectra falling off much slower than the anticipated exponential trend~\cite{cronin10,cronin0}. The mechanism(s) for transport of a major fraction of projectile momentum to transverse momentum space far from projectile rapidity were sought theoretically in the context of newly-formulated perturbative QCD (pQCD).

As a complementary effort various empirical spectrum models have been proposed. With initial publication of its S$ p\bar p$S data~\cite{ua1power} the UA1 collaboration (1982) proposed an empirical function
\bea \label{ua1pow}
E\frac{d^3\sigma}{dp^3} &\propto& \frac{p_{t_0}^n}{(p_{t_0} + p_t )^n} \leftrightarrow \frac{1}{(1+p_t/p_{t_0})^n},
\eea
sometimes referred to as a power-law or Hagedorn function, which visually appears to transition from an exponential at lower \pt\ to a power law at higher \pt~\cite{hagedorn}. One theoretical response to observed spectrum shape evolution with event multiplicity was the suggestion that a quark-gluon plasma or QGP is formed at such high collision energies~\cite{vanhove,hagedorn}. In Ref.~\cite{pancheriqgp} the QGP conjecture is acknowledged along with one from Ref.~\cite{jacob}:  ``...an energy threshold has been crossed such that many low-$x$ partons have now enough energy to undergo hard scattering and produce several mini-jets, of a few GeV each, which fragment independently from one another.''  Reference~\cite{pancheriqgp} concludes: ``...the multiplicity dependence of inclusive \pt\ spectr[um shapes] reported by UA1 can probably be explained by invoking the persistent high multiplicity associated with low energy (few GeV) jet events.'' 

That concept was followed up by theoretical proposal of a {\em two-component model} of hadron production in high-energy collisions~\cite{pancheri} (1985) of the form
\bea \label{lia}
\frac{d\sigma}{dndp_t} &=& \frac{d\sigma_0}{dn} P_{0}(p_t) +  \frac{d\sigma_{1}}{dn} P_{1}(p_t),
\eea
where the first term applies to soft or nonjet processes and the second represents a contribution from ``low-$x$ QCD jets'' or so-called minijets. $d\sigma_x/dn$ represent cross section fractions associated with event multiplicity $n$, and $P_x(p_t)$ are unit-normal model functions. That theoretical model is then formally equivalent to TCM Eq.~(\ref{tcmspec}) below.

An earlier empirical model was introduced (1978) to describe dimuon spectra from $p$-A collisions~\cite{yohqgauss} (FNAL) with the form
\bea \label{plold}
E\frac{d^3\sigma}{dp^3} &\propto&  \frac{1}{[1+(p_t/p_{t_0})^2]^n}.
\eea
The same model has since been used to describe higher-mass hadron spectra~\cite{alijpsi,phenixqgauss}. That model is formally equivalent to the so-called power-law element of the Bylinkin model discussed below.

More recently, a spectrum model intended to describe thermodynamic systems with incomplete equilibration, associated with so-called Tsallis statistics~\cite{tsallis} (1988), has become popular
\bea \label{tsallis}
\frac{d^2\sigma}{p_t dp_tdy_z} &\propto& \frac{1}{(1+m_t/T n)^n},
\eea
where $m_t = \sqrt{p_t^2 + m_0^2}$ is transverse mass, $T$ is interpreted as a thermodynamic temperature and $1/n \leftrightarrow q-1$ is interpreted as a measure of incomplete equilibration. In a nuclear physics context $1/n > 0$ may be attributed to hard processes depending on jet parameters. The formal similarity between Eq.~(\ref{tsallis}) and Eq.~(\ref{ua1pow}) is notable. 

A comment on notation: The TCM hard-component exponential tail is determined by parameter $q$ as is the exponent of the Tsallis function above. In this text parameter $q$ is reserved for TCM hard component $\hat H_0(y_t)$ and exponent $n$ for soft component $\hat S_0(m_t)$. Transverse rapidity is defined by $y_t = \ln((m_t + p_t)/m_0)$. Symbol $N$ is reserved for the Bylinkin power-law element while $n$ is also used for miscellaneous exponents such as in Eq.~(\ref{ua1pow}).

Single-element models such as Eqs.~(\ref{ua1pow}), (\ref{plold}) and (\ref{tsallis}) can be described as {\em monolithic} and may be based on  {\em a priori} assumptions including a thermalization process (via multiple scattering) as central to HE collisions. As such, monolithic models do not share the attributes of the TCM proposed theoretically in Ref.~\cite{pancheri} wherein separate jet and nonjet hadron {\em production mechanisms} are acknowledged and addressed by distinct approaches.

\subsection{Empirical two-component model or TCM}

As an alternative to experimental or theoretical {\em a priori} conjecture one may analyze spectrum data systematics in an attempt to identify what mathematical model is {\em required} by data. That process can be viewed as  a form of lossless data compression~\cite{tommodeltests}: What is the minimum representation (in terms of algebraic structure and parameter number) that can fully represent all information in particle data to the limits of statistical uncertainty?

Reference~\cite{ppprd} (2006) describes such a procedure, having as its central goal an exhaustive study of the evolution of spectrum shapes with event multiplicity \nch. There are no {\em a priori} assumptions about spectrum structure or theoretical interpretation. The study has two principal outcomes: (a) spectra are composed of two distinct parts with different functional forms independent of event multiplicity \nch\ or nearly so, and (b) the particle yield associated with one part {\em varies as the square} of the yield for the other part. The second property provides an accurate method to separate the parts and characterize their functional forms. That approach is  consistent with Eq.~(\ref{lia}).

\begin{figure}[h]
\includegraphics[height=1.65in]{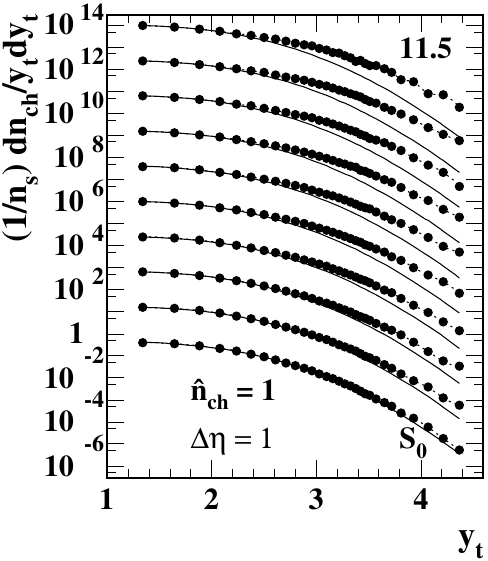}
\includegraphics[height=1.65in]{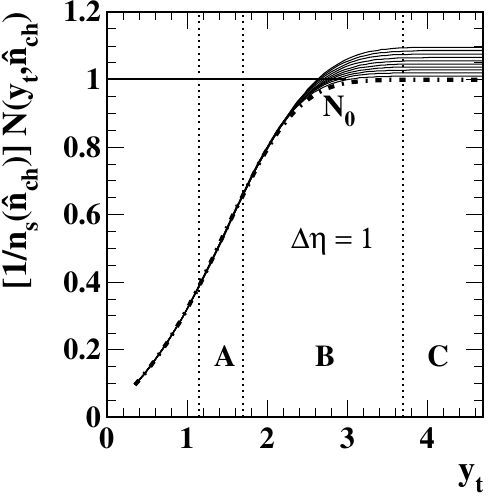}
\caption{\label{pp1}
Left: \yt\ spectra from ten charge-multiplicity classes of 200 GeV \pp\ collisions (points) compared to fixed reference $\hat S_0(y_t)$ (solid curves)~\cite{ppprd}.
Right: Running integrals $N(y_t)$ of normalized (and extrapolated) $y_t$ spectra in the left panel for ten multiplicity classes (solid curves) compared to running integral $N_0(y_t)$ of fixed reference $\hat S_0(y_t)$ (dash-dotted curve).  $\hat n_{ch} \approx n_{ch} / 2$ is uncorrected event multiplicity. \yt\ = 2 corresponds to $p_t \approx 0.5$ GeV/c and \yt\ = 3.75 to $p_t \approx 3$ GeV/c
}  % ppcomm14bb, integrals2xxx 
\end{figure}

Figure~\ref{pp1} provides an illustration from Ref.~\cite{ppprd}. The left panel shows \yt\ spectra (points) for ten \pp\ multiplicity classes. Empirically, all data spectra are observed to coincide at lower \yt\ if rescaled by  soft-component multiplicity $n_s \approx n_{ch} - \alpha n_{ch}^2$ for some $\alpha \approx 0.01$. As plotted the spectra are displaced upward from each other by successive factors 40 relative to the lowest spectrum

%%%%%%%%%%%%%%%%%%%%%%%%%%%%%%%%%%
%%%%%%%%%%%%%%%%%%%%%%%%%%%%%%%%%%

Figure~\ref{pp1} (right) shows {\em running integrals} of the ten rescaled spectra in the left panel. A running integral may enhance a signal (spectrum shape) over statistical noise. $N_0(y_t)$ is an asymptotic limit. Different behavior is observed in each of \yt\ intervals A, B and C. In interval C the integrals are nearly constant, and those constant values increase approximately linearly with $n_s$ relative to $N_0$.  The main {\em shape variation} lies within interval B, effectively as the running integral of a {\em peaked spectrum component} (hard component) with amplitude $\propto n_{s}$.

The resulting TCM has been applied to many collision systems and to both unidentified hadrons~\cite{ppprd,alicetommpt,alicetomspec,tommpt} and identified hadrons~\cite{ppbpid,pidpart1,pidpart2,transport}. As a result, a {\em self-consistent system} of simple parameter variation with \mbox{A-B} collision system, hadron species and collision energy has been established. The TCM is thereby a {\em formally predictive} model (as opposed to theoretical prediction). With the algebraic properties of the two components determined,  corresponding physical mechanisms have been established {\em a posteriori}: projectile hadron dissociation (soft component) and low-$x$ gluon large-angle scattering and fragmentation to jets (minijets, hard component).

\subsection{Bylinkin two-element model}

Subsequent to Ref.~\cite{ppprd} (2006) an alternative ``two-component'' model has been proposed in Ref.~\cite{bylinkin2010} (2010) consisting of an exponential element and a ``power-law'' element, where the quotes question the terminology
\bea \label{byeq}
\frac{d\sigma}{p_t dp_t} &=& A_{th} \exp[-(m_t - m_0) / T_{th}] + \frac{A_h}{\left(1 + \frac{p_t^2}{T_h^2 N}\right)^N}.~~~
\eea
That model is described in Ref.~\cite{bylinkinanom} as  ``a modified Tsallis-type function.'' The two elements are described in Ref.~\cite{bylinkinprd}  thus: ``The power-like component [second element] is mainly originated from a relatively large \pt\ domain, that is from the mini-jet fragmentation, while the exponential part accounts for some `thermalization' caused by the final state rescattering in the parton cloud and the `hadron gas' formed by the secondaries.''

There are several issues with that model before data are encountered. In this text an ``element'' of a model is distinguished from a ``component'' of physical hadron production with corresponding model representation. The second element of Eq.~(\ref{byeq}) is, in a Tsallis context, a $q$-Gaussian with low-\pt\ limiting form  $\exp(-p_t^2/T_h^2)$ centered at zero \pt\ and high-\pt\ power-law tail $\propto p_t^{-2N}$. The relation between this ``power-law'' term and Eq.~(\ref{plold}) employed to describe entire spectra is notable. The Bylinkin power-law element straddles two production mechanisms modeled as distinct by the TCM described in the previous subsection -- projectile dissociation and jet production. 

In material below it will be established that several conclusions associated with the Bylinkin model conflict with data properties.  Among them: the low-\pt\ part of a spectrum is described by an exponential that represents thermalization, possibly by an exotic process; the thermalized portion is present for pions but not other hadrons; a jet contribution is described by a $q$-Gaussian; two ``temperatures'' characterizing those elements are linearly coupled; and certain data statistics vs event multiplicity and collision energy follow power-law trends.

%%%%%%%%%%%%%%%%
\section{Data-model-model comparisons} \label{compare}

In  this section several issues relating to the Bylinkin model are discussed in comparison with various data formats and the TCM. They include evolution of model shapes with hadron mass, evolution of power-law parameters with collision energy and event \nch, evolution of soft and hard charge densities with event \nch\ and collision energy, and variation of ensemble-mean \mmpt\ inferred from the two models with \nch\ and collision energy.

\subsection{Bylinkin model properties vs TCM}

This subsection responds to descriptions of the Bylinkin model in Refs.~\cite{bylinkin2010,bylinkinanom,bylinkphen}. Several assumptions associated with the Bylinkin model are notable: It is observed that with increasing \pt, spectra transition from exponential (Boltzmann-like) to power law ``traditionally interpreted as an onset of the perturbative QCD regime of hadron production. ... These features...are found to be {\em universal for any type of colliding particles} [emphasis added]. Therefore, it is tempting to find one universal smooth functional form...'' to model \pt\ spectra.  ``...the parameter $T_h$ is a QCD analogy to a temperature in classical thermodynamics.'' ``...the physical origin of the observed $T_{th}$-$T_h$ correlation is not quite clear....''

The model elements are interpreted thusly: ``As required by the perturbative QCD the spectrum of particles produced in parton-parton interactions is described by a power law distribution. The rest bulk of irradiated hadrons represents a quasi-thermolized [sic] hadronic gas produced with a characteristic temperature $T_{th}$. The spectrum of hadrons in this gas has the Boltzman [sic] exponential shape. ... only the inclusive spectra of charged particles produced in pure baryonic collisions require a substantial contribution  of the Boltzman-like [sic] exponential term.'' ``The difference in size [i.e.\ relative amplitude] of the exponential [element]...is mainly responsible for the difference in the spectra shapes...'' \cite{bylinkin2010}.

In  the several papers presenting the Bylinkin model as applied to spectrum data the plotted spectra are almost always minimum-bias (MB). Such spectra have a minimal jet contribution (i.e.\ are dominated by the soft component), and challenges from substantial spectrum evolution with event multiplicity \nch\ are not addressed.

\begin{figure}[h]
	\includegraphics[height=1.6in]{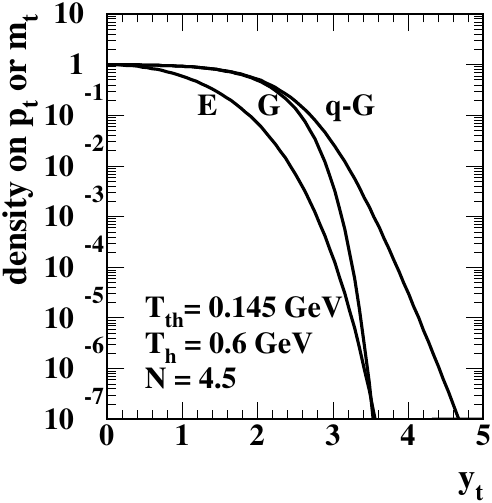}
	\includegraphics[height=1.6in]{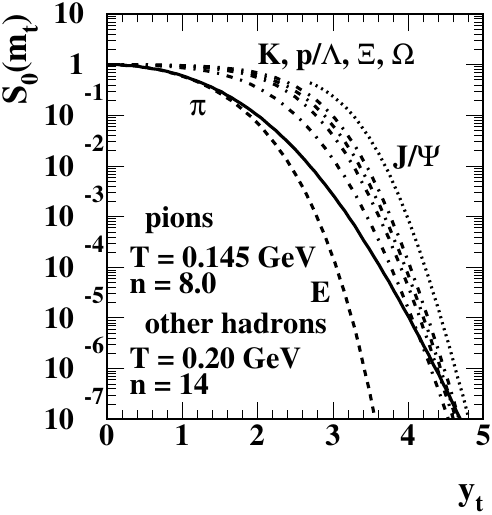}
	\caption{\label{shape}
Left: Bylinkin model elements: E is the exponential element, $q$-G the power-law element ($q$-Gaussian) and G its asymptotic limit.
Right: TCM soft component $\hat S_0(m_{ti})$ for several  species from pions to J$/\Psi$ illustrating mass dependence. E is the exponential  limit of the pion soft component.
		}   % acompare1, 2
\end{figure}

Figure.~\ref{shape} (left) shows the two elements of the Bylinkin model presented as curves E (exponential) and $q$-G (power law) vs pion transverse rapidity \yt. The parameter values are taken from Ref.~\cite{bylinkin2010} Figs.~3 and 8 (central values). As noted, the nominal power-law element  has the form of a $q$-Gaussian -- a Gaussian with power-law tail. The curve labeled G is the Gaussian limiting case with $1/N \rightarrow 0$. Since the ``power-law'' term is associated with jet production this model asserts that jet fragments extend down to zero \pt\ and are {\em maximum} at that point, conflicting with the popular notion that jet production is a ``high \pt'' phenomenon. The $T_h$ parameter controls the position on \yt\ of the apparent shoulder of the model, and $N$ controls the slope of the high-\pt\ tail.

Referring to its Fig.~3 Ref.~\cite{bylinkin2010} comments ``The most surprising feature of the new parameterization (2) is a strong [linear] correlation between the parameters $T_{th}$ and $T_h$.'' But that is not surprising if the two model elements strongly overlap at lower \pt\ and neither one is well suited to model the targeted hadron production process. In essence,  the two model elements have a conflict of interest that is clearly demonstrated by the trivial relation between two parameters. In contrast,  soft and hard model parameters of the TCM have unique properties clearly related to their respective targeted spectrum components. Note changes $n \rightarrow N$, $T_e \rightarrow T_{th}$ and $T \rightarrow T_h$ have been made in this text to minimize confusion.

%%%%%%%%%%
%%%%%%%%%%%%

Figure~\ref{shape} (right) shows TCM soft component $\hat S_0(m_{ti})$ for several hadron species $i$ of increasing mass from pions to $J/\Psi$. Dashed curve E is the exponential limit of pion $\hat S_0(m_t)$, again for $1/n \rightarrow 0$. The $\hat S_0(m_{ti})$ mass dependence has two consequences: (a) the shoulder moves to higher \pt\ leaving a nearly-flat interval below it and (b) the slope at higher \pt\ increases (i.e.\ the effective slope parameter $T$ decreases). Those trends can both be understood by the relation $(m_{ti} - m_{0i})/T = m_{0i}[\cosh(y_{ti}) - 1] / T$. Property (a) arises from the $\cosh(y_{ti})$ term where $y_{ti} = \ln[(m_{ti} + p_t)/m_{0i}]$ is the proper transverse rapidity for species $i$ {\em including its mass dependence}. Property (b) arises because the effective slope parameter is $T/m_{0i}$.

The Bylinkin model functions have no such mass dependence. To accommodate spectra for different hadron species and collision conditions  model parameters $T_{th}$, $T_h$ and $N$ plus amplitudes $A_{th}$ and $A_h$ must vary substantially. For more-massive hadrons the flat interval below the shoulder, increasing in width with hadron mass, reduces {\em or excludes} the Bylinkin exponential term as noted in Refs.~\cite{bylinkinanom,bylinkphen}. Parameter $T_h$, effectively the width $\sigma$ of the $q$-Gaussian, attempts to follow shifts of the  $\hat S_0(m_{ti})$ shoulder to higher \pt, {\em and} growth of the jet-related hard component also at higher \pt, with increasing event \nch. If the exponential term is ruled out for baryons and some collision systems as reported in Refs.~\cite{bylinkinanom,bylinkintherm} Bylinkin parameters $T_h$ and $N$ alone must accommodate both $\hat S_0(m_{ti})$ evolving with hadron mass and $\hat H_0(y_t)$ evolving with hadron species, collision system and energy.

\subsection{Power-law element parameter evolution}

This subsection responds to descriptions of Bylinkin model parameter variations in Refs.~\cite{bylinkin2010,bylinkin2015,bylinkphen}. Bylinkin parameter trends are compared directly with those from TCM analysis of \pp\ spectra at several collision energies.

\begin{figure}[h]
	\includegraphics[height=1.6in]{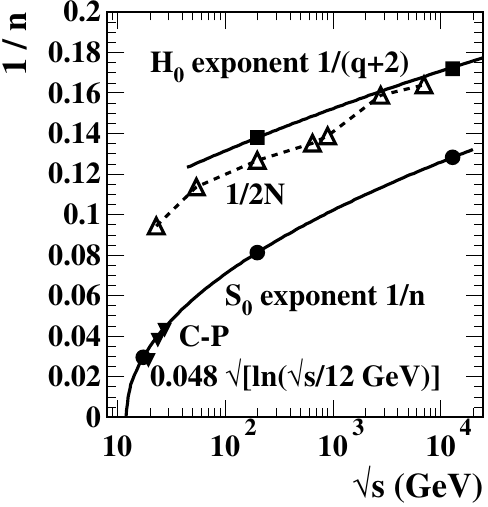}
	\includegraphics[height=1.6in]{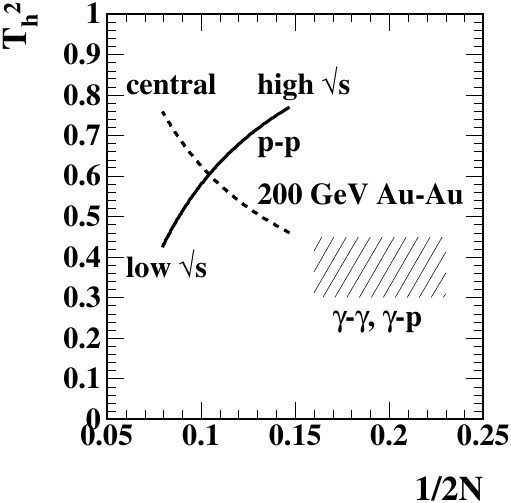}
	\caption{\label{power}
Left: Bylinkin power-law exponent $N$ (open triangles) from Ref.~\cite{bylinkin2015} compared to TCM soft-component exponent $n$ (lower solid curve and points) and hard-component exponent $q$ (upper solid curve and points) from Ref.~\cite{alicetomspec} vs collision energy.
Right: Correlated evolution of Bylinkin power-law parameters $T$ and $N$ for \pp\ collisions vs collision energy (solid) and \auau\ collisions vs centrality (dashed) from Ref.~\cite{bylinkin2010}.
		} %  alice140bxby,  acompare8
\end{figure}

Figure~\ref{power} (left) shows Bylinkin power-law exponent $N$ (open triangles) from Fig.~3 of Ref.~\cite{bylinkin2015} plotted as $1/2N$. Those values are compared to TCM hard-component exponent $q$ plotted as $1/(q+2)$ (squares) and TCM soft-component exponent in the form $1/n$ (dots, triangles) from Ref.~\cite{alicetomspec}. This plot makes clear that the Bylinkin power-law model approximates by its tail the exponential tail (on \yt) of the TCM hard component and is quite different from the power-law tail of the TCM soft component. The combination $q + 2$ (with $q$ on \yt) approximates the power law (on \pt) of the TCM hard-component tail.

Referring to Ref.~\cite{bylinkin2015} Fig.~3, decrease of $N$ with $\sqrt{s}$ (i.e.\ increase of $1/2N$) is ``...related to the fact that the probability to produce a high-\pt\ mini-jet should grow with $\sqrt{s}$.'' Actually, it is likely not the {\em rate} of jet production (probability to produce) but the {\em shape} of the spectrum hard component (``hardening'') that changes with $\sqrt{s}$ reflecting evolution of the underlying jet energy spectrum with increasing \pp\ collision energy~\cite{fragevo,jetspec2}.

Referring to Fig.~2 of Ref.~\cite{bylinkin2015} ``Note that the parameter [$N$] shows a growth [decrease of $1/2N$] with pseudorapidity that is explained by higher thermalization of the spectra....'' More likely, the trend is due to the dramatic difference between soft- and hard-component $\eta$ distributions such that at higher $\eta$ spectra are dominated by the soft component (see Fig.~\ref{etaquad}). From Ref.~\cite{bylinkin2015} Fig.~2 $\eta = 0$ corresponds to $1/2N \approx 0.14$ whereas for $\eta = 3$ $1/2N \approx 0.11$, then see the TCM values from solid curves at left corresponding to $\sqrt{s} = 630$ GeV per Ref.~\cite{bocquet}.

The energy trend for TCM $\hat S_0(m_t)$ exponent $n$ in the left panel appears consistent with the concept of {\em Gribov diffusion}~\cite{gribov,gribov3,gribov2} such that within a parton splitting cascade there is random walk of parton \pt\ from generation to generation leading to hardening of the soft-component tail with increased $\sqrt{s}$ or depth of the splitting cascade.

%%%%%%%%%%
%%%%%%%%%%%%

Figure~\ref{power} (right) shows Bylinkin power-law $T_h^2$ vs $N$ trends from Fig.~8 of Ref.~\cite{bylinkin2010} for \pp\ collisions (solid) varying with collision energy and 200 GeV \auau\ collisions (dashed) varying with centrality. The power-law exponent is plotted as $1/2N$ to be compatible with the TCM hard-component power-law tail on \pt\ with effective exponent $1/(q+2)$. Quantity 1/exponent measures the slope of the power law. The hatched region represents $\gamma$-$\gamma$ and $\gamma$-$p$ results from the same figure in Ref.~\cite{bylinkin2010}. 

Those trends can be simply interpreted: The \pp\ trend corresponds to a large increase in the jet contribution to \pp\ spectra with substantial hardening of the distribution tail as demonstrated in Fig.~3 (left) of Ref.~\cite{tommpt}. Thus, Bylinkin parameters $T_h^2$ and $1/2N$ must both increase together substantially. The peripheral end (bottom) of the \auau\ curve should correspond to 200 GeV \pp, and $1/2N$ is thus expected to have a value near 0.13 per the left panel. One expects a large increase in $T_h^2$ with \auau\ centrality, but the corresponding {\em decrease} in $1/2N$ is consistent with high-\pt\ suppression associated with jet quenching. The hatched band indicates that $\gamma$-$\gamma$ and \mbox{$\gamma$-$p$} spectra are modeled by a $q$-Gaussian with small width and hard power-law tail. Compare with Fig.~\ref{ggmm}.

\subsection{Particle densities $\bf \bar \rho_{x}$ vs p-p event $\bf n_{ch}$ and $\bf \sqrt{s}$}

This subsection responds to particle density variation with changing event conditions inferred from Bylinkin model fits to \pp\ \pt\ spectra. Reference~\cite{bylinkinprd} reports particle densities vs event multiplicity \nch\ and collision energy $\sqrt{s}$. For both \nch\ and $\sqrt{s}$ trends  fits to spectrum data with the Bylinkin model return soft and hard particle densities $\bar \rho_{sb}$ and $\bar \rho_{hb}$. The \nch\ trend is derived from hadron spectra for 200 GeV \pp\ collisions presented in Ref.~\cite{ppprd}.  It is notable that spectrum data from Ref.~\cite{ppprd} are employed but there is no acknowledgment of {\em first detailed derivation} of the TCM reported therein. 

Regarding particle-density dependence on event \nch\ Ref.~\cite{bylinkinprd} asserts that ``One can notice that the contribution from the power-like component (mini-jets) [$\bar \rho_{hb}$] grows faster than that from the `exponential' one [$\bar \rho_{sb}$]. ... This fact confirms that partly the increase of $N_{ch}$ is due to a larger $E_t$ of mini-jets produced.'' To the contrary, in Ref.~\cite{ppprd} it is established that the \pp\ spectrum hard component {\em shape}, which reflects MB jet energy~\cite{fragevo,jetspec2}, is independent of event \nch\ to good approximation. The relative rates of increase with \nch\ reflect a {\em quadratic relation} between hard and soft particle densities which is a central discovery of Ref.~\cite{ppprd}. It is thus informative to compare Bylinkin model trends with TCM results.

\begin{figure}[h]
	\includegraphics[height=1.6in]{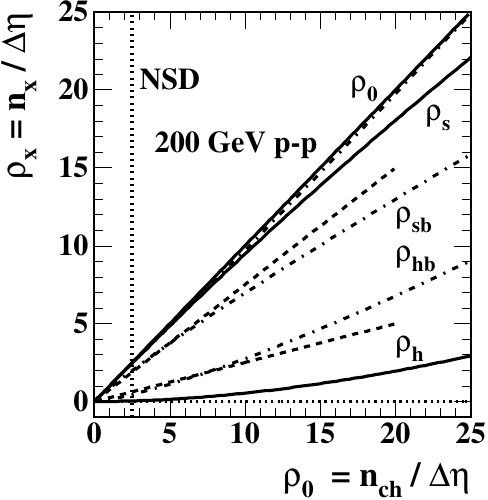}
	\includegraphics[height=1.6in]{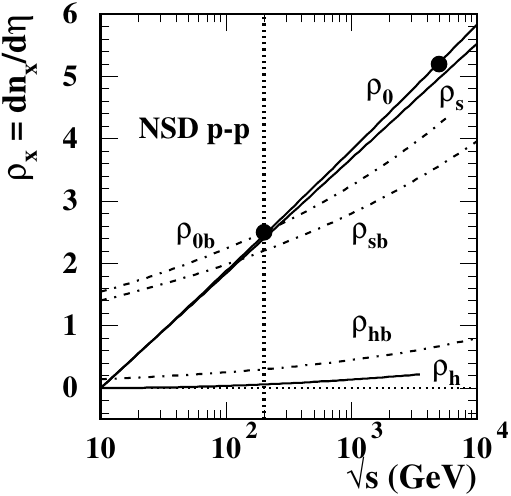}
	\caption{\label{charge}
Left: Bylinkin exponential $\bar \rho_{sb}$ and power-law $\bar \rho_{hb}$ particle densities (dash-dotted) vs total particle density $\bar \rho_{0}$~\cite{bylinkinprd} compared to equivalent trends for TCM $\bar \rho_{s}$ and hard $\bar \rho_{h}$ (solid). The dashed lines are linear extrapolations of Bylinkin  trends at lower densities.
Right: Bylinkin (dash-dotted) and TCM (solid) particle densities vs collision energy $\sqrt{s}$. The solid dots represent NSD $\bar \rho_0$ for 200 GeV and 5 TeV. The dotted line represents the collision energy for the left panel.
		}  % acompare3, 6
\end{figure}

Figure~\ref{charge} (left) shows particle densities $\bar \rho_{sb}$ and $\bar \rho_{hb}$ (dash-dotted)  inferred with the Bylinkin model vs {\em corrected} event multiplicity \nch. The same trends presented in Fig.~5 of Ref.~\cite{bylinkinprd} include quantity $N_{ch}$ which is actually the {\em uncorrected} $\hat n_{ch}$ from Ref.~\cite{ppprd} where $n_{ch} \approx 2\hat n_{ch}$. Those trends can be compared with TCM trends  $\bar \rho_{s}$ and $\bar \rho_{h}$ (solid) predicted by the TCM relation $\bar \rho_h \approx \alpha \bar \rho_s^2$ (for \pp\ collisions) with $\alpha \approx 0.006$ for 200 GeV~\cite{ppprd}. Their sum $\bar \rho_0$ is indicated by the uppermost solid line. The dashed lines extrapolate linearly the $\bar \rho_{sb}$ and $\bar \rho_{hb}$ trends for low \nch\ and represent 25\% and 75\% of the total represented by the Bylinkin sum (uppermost dash-dotted). This result suggests that the Bylinkin power-law element includes 1/4 of the actual data soft component as a bias. Relative to those linear trends the curvatures of $\bar \rho_{sb}$ and $\bar \rho_{hb}$ do approximately match the TCM versions.

%%%%%%%%%%
%%%%%%%%%%%%

Figure~\ref{charge} (right)  shows Bylinkin-model charge densities $\bar \rho_{sb}$ and $\bar \rho_{hb}$ (dash-dotted) vs collision energy $\sqrt{s}$  as reported in Fig.~2 of Ref.~\cite{bylinkinprd}. The curves represent
\bea
\bar \rho_{sb}(n_{ch},s) &=& \bar \rho_{sb}(n_{ch})(s/s_0)^{0.15}
\\ \nonumber
\bar \rho_{hb}(n_{ch},s) &=& \bar \rho_{hb}(n_{ch})(s/s_0)^{0.25}
\eea
repeating Eqs.~(12) and (13) of Ref.~\cite{bylinkinprd},
with $\sqrt{s_0} = 200$ GeV and $\bar \rho_{xb}(n_{ch})$ corresponding to NSD values 0.3 and 2.1 in the left panel. The TCM predictions (solid) include the trend $\bar \rho_s \approx 0.8 \ln(\sqrt{s} / 10~ \text{GeV})$~\cite{alicetomspec} and $\bar \rho_h \approx \alpha(\sqrt{s}) \bar \rho_s^2$ with $\alpha(\sqrt{s}) \in [0.006,0.017] \approx 0.01$ here for simplicity. The TCM trend $\bar \rho_0(\sqrt{s})$ provides a good description of NSD \pp\ values from 200 GeV to 13 TeV. Solid dots represent NSD $\bar \rho_0$ values for 200 GeV and 5 TeV. Note that this $\bar \rho_s$ trend for {\em gluon} production does not include substantial quark-related production at lower energies.

\subsection{Mean values $\bf \langle p_{t}\rangle_x$ vs p-p event $\bf n_{ch}$ and $\bf \sqrt{s}$} \label{mpttrends}

This subsection responds to ensemble-mean \mmpt\ variation on event multiplicity \nch\ and collision energy $\sqrt{s}$ inferred from fits to \pp\ \pt\ spectra with the Bylinkin model as reported in Ref.~\cite{bylinkinprd}. Those results are compared to corresponding trends from the TCM.

\begin{figure}[h]
	\includegraphics[height=1.6in]{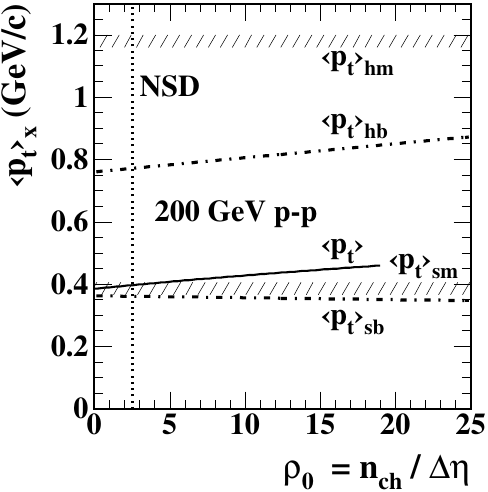}
	\includegraphics[height=1.6in]{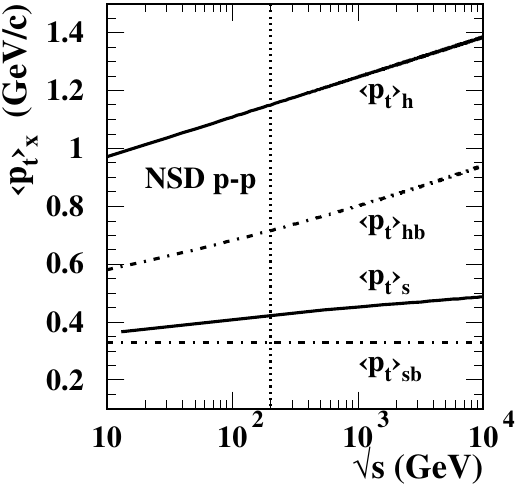}
	\caption{\label{mptby}
Left: Bylinkin exponential $\langle p_t \rangle_{sb}$ and power-law $\langle p_t \rangle_{hb}$ means vs total particle density $\bar \rho_{0}$ (dash-dotted)~\cite{bylinkinprd} compared to  trends  $\langle p_t \rangle_{xm}$ for TCM soft- and hard-component model functions (hatched bands). The solid curve is the combined TCM \mpt\ prediction.
Right:  Bylinkin exponential $\langle p_t \rangle_{sb}$ and power-law $\langle p_t \rangle_{hb}$ means vs collision energy $\sqrt{s}$ (dash-dotted)~\cite{bylinkinprd} compared to TCM trends (solid).
		}  %  acompare4, 7
\end{figure}

Figure~\ref{mptby} (left) shows ensemble-mean contributions \mpt$_{xb}$ (dash-dotted lines) inferred from the two elements of the Bylinkin spectrum model as reported in Fig.~4 of Ref.~\cite{bylinkinprd} for 200 GeV \pp\ collision data taken from Ref. \cite{ppprd}. Also shown are \mpt$_{xm}\leftrightarrow \bar p_{txm}$ trends for soft and hard components of the corresponding TCM (lower and upper dashed bands) and total ensemble-mean \mpt\ (solid curve) as reported in Ref.~\cite{ppprd}. The last is evaluated as 
\bea
\langle p_t \rangle &=& \frac{\bar \rho_s \bar p_{tsm} + \bar \rho_h \bar p_{thm}}{\bar \rho_s + \bar \rho_h }
\eea
with the $\bar p_{txm}$ derived from invariant TCM model functions. Densities $\bar \rho_s$ and $\bar \rho_h$ are as shown in Fig.~\ref{charge} (left).

The Bylinkin model result for \mpt$_{hb}$ can be explained in terms of the ``power-law'' element from which it is derived. Because the latter  is essentially a $q$-Gaussian with centroid at zero \pt\ the inferred mean values must be well below $\bar p_{thm}$ for the TCM hard component, a Gaussian on \yt\ with mode near 1 GeV/c. The small increase of \mpt$_{hb}$ is due to variation of Bylinkin parameter $T_h$ ($q$-Gaussian width) with \nch\ as in Fig.~3 of Ref.~\cite{bylinkin2010}, presumably in response to {\em relative} increase of jet production.

%%%%%%%%%%
%%%%%%%%%%%%

Figure~\ref{mptby} (right) shows collision-energy $\sqrt{s}$ dependence of soft and hard \mpt\ components (dash-dotted) from the Bylinkin model as presented in Fig.~3 of Ref.~\cite{bylinkinprd}. The \mpt$_{hb}$ trend is reported as $\propto (\sqrt{s})^{0.07}$. Also presented are \mpt$_{x}$ trends from the TCM, with the \mpt$_{h}$ trend (solid line) taken from  Fig.~3 (right) of \cite{tommpt} and the \mpt$_{s}$ trend based on the usual TCM $\hat S_0(m_t)$ for pions and the exponent $1/n(\sqrt{s})$ energy trend from Fig.~\ref{power} (left). The resulting values for \mpt$_{s}$ are slightly high because no model for the low-\pt\ resonance contribution is included. Reference \cite{tommpt} explains the rise in \mpt$_{h}$ with $\sqrt{s}$ as due to corresponding evolution of the spectrum hard component (jet fragment distribution) resulting from variation of the underlying jet energy spectrum with collision energy.

%%%%%%%%%%%%
\section{Models and data compression} \label{compress}

This section describes differential methods to obtain optimized models for density distributions on \pt\ and $\eta$ that represent {\em lossless data compression}: the input data are reduced to a minimal set of functional forms and parameters that retain almost all information carried by the data but do not rely on {\em a priori} assumptions. A further constraint is that the functional forms and their parameters must be {\em a posteriori interpretable}, that is, directly and quantitatively comparable with QCD theory.

\subsection{Spectrum TCM with (almost) no assumptions} \label{subtract}

This subsection briefly summarizes the process of inferring TCM model functions from 200 GeV \pp\ \pt\ spectra. Reference~\cite{ppprd} (2006) reported a two-component model of \pt\ spectra from 200 GeV \pp\ collisions. The model structure, inferred empirically from spectrum data, is represented by
\bea \label{tcmspec}
\bar \rho_{0}(p_t,n_s) &\approx&  d^2 n_{ch} / p_t dp_t d\eta
\\ \nonumber
&=& S(p_t,n_s) + H(p_t,n_s)
\\ \nonumber
&=&  \bar \rho_{s} \hat S_{0}(p_t) +  \bar \rho_{h} \hat H_{0}(p_t,n_s),
\eea
where $\bar \rho_x = n_x / \Delta \eta$ within $\eta$ acceptance $\Delta \eta$ and $n_s$ is used as an event index. The main feature is two distinct components (hadron production processes) each of which {\em factorizes} charge density and \pt\ dependence. Of central importance is discovery of  the relation $\bar \rho_h \propto \bar \rho_s^2$. Based on that system it is possible to infer the functional forms of $\hat S_{0}(p_t)$ and $\hat H_{0}(p_t,n_s)$ by simple difference methods.

\begin{figure}[h]
	\includegraphics[height=1.65in]{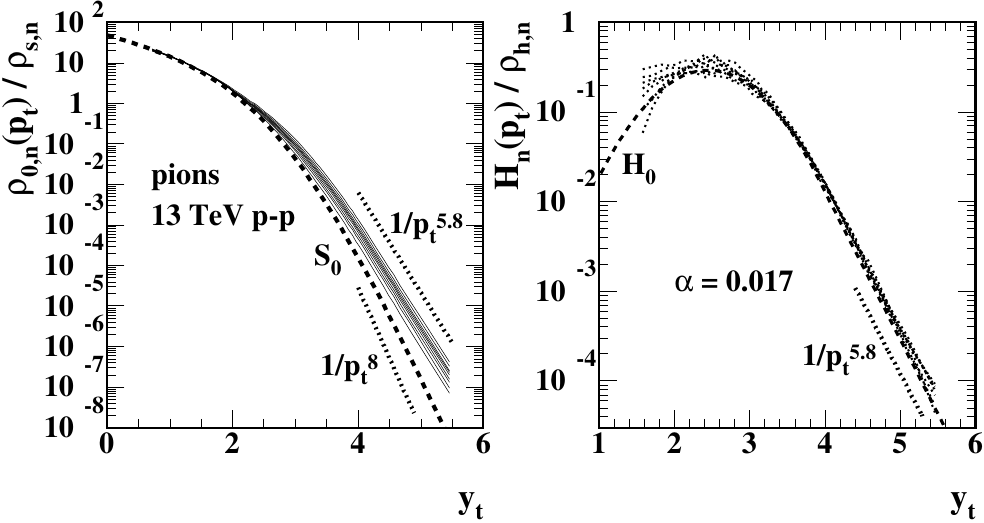}
	\caption{\label{none}
Left: Pion spectra for ten event classes from 13 TeV \pp\ collisions (solid) rescaled to the form $\bar \rho_0(m_t,n_s)/\bar \rho_s$ vs pion transverse rapidity \yt~\cite{pppid}.  The dashed curve is soft component $ S_0'(m_t)$ defined by Eq.~(\ref{s0prime}).
Right: Successive spectrum differences defined by Eq.~(\ref{etahcx}) revealing data hard components.  The dashed curve is TCM hard-component model $\hat H_0(y_t)$. Hard-component data and model are densities on \yt.
		}   % acompare10
\end{figure}

Figure~\ref{none} (left) shows spectra $\bar \rho_{0}(p_t)$ as densities on \pt\ plotted vs pion transverse rapidity \yt\ for ten event classes $n$ from 13 TeV \pp\ collisions. Spectra have been rescaled to the form $\bar \rho_{0}(p_t) / \bar \rho_{s}$ where soft particle density $\bar \rho_{s}$ is derived from measured charge density $\bar \rho_{0}$ as the root of the relation $\bar \rho_{0} \approx \bar \rho_{s} + \alpha(\sqrt{s}) \bar \rho_{s}^2$, and $\alpha(\sqrt{s}) \approx 0.017$ has been determined previously from systematic analysis of several collision systems. See Fig.~16 (left) of Ref.~\cite{alicetomspec}.

%%%%%%%%%%
%%%%%%%%%%%%

Figure~\ref{none} (right) shows normalized data hard components $H_n(p_t)$ (dotted curves) derived from full spectra by the following procedure. Based on Eq.~(\ref{tcmspec}) the ratio $\bar \rho_{0,n}(p_t) / \bar \rho_{s,n} \approx \hat S_{0}(p_t) + H_{n}(p_t)/\bar \rho_{s,n}$, where $H_{n}(p_t)$ here represents a {\em data} hard component. If ratio $H_{n}(p_t)/\bar \rho_{h,n}$ is invariant or slowly varying with event class $n$ (as observed for various systems) it can be inferred from rescaled spectra by the following difference procedure
\bea \label{etahcx}
H_{n}(p_t) / \bar \rho_{h,n} &\approx& \frac{\bar \rho_{0,n}(p_t) / \bar \rho_{s,n} - \bar \rho_{0,n-1}(p_t) / \bar \rho_{s,n-1}} {\bar \rho_{h,n} / \bar \rho_{s,n} - \bar \rho_{h,n-1} /   \bar \rho_{s,n-1}},
\eea
where $\hat S_{0}(p_t)$ cancels in the numerator. By assumption $ H_{n}(p_t) / \bar \rho_{h,n} \approx H_{n-1}(p_t) / \bar \rho_{h,n-1}$ and  factors out of the numerator difference. The result is dotted curves in the right panel confirming assumed  invariance of $H_{n}(p_t)/\bar \rho_{h,n}$. The corresponding TCM hard-component fixed model $\hat H_{0}(p_t)$ is shown by the dashed curve. The two inferred TCM model functions are defined by
\bea \label{s00}
\hat S_{0i}(m_{ti}) &=& \frac{A_i}{[1 + (m_{ti} - m_i) / T_i n_i]^{n_i}},
\eea
as density on \mt\ for hadron species $i$ and
\bea \label{h00}
\hat H_{0}(y_t) &\approx & B \exp\left\{ - \frac{(y_t - \bar y_t)^2}{2 \sigma^2_{y_t}}\right\}~~\text{near mode $\bar y_t$}
\\ \nonumber
&\propto &  \exp(- q y_t)~~~\text{for higher $y_t$ -- the tail}
\eea
 as density on \yt\ converted to \mt\ by Jacobian $y_t / m_t p_t$.

It should be emphasized that the only untested (in this procedure) assumption is the value of $\alpha$ for 13 TeV \pp\ collisions that has been derived from separate studies~\cite{alicetomspec}. The single parameter value permits identification of $ \hat S_{0}(m_t)$ as the dashed curve in the left panel, {\em the asymptotic limit} of rescaled data spectra for $n_{ch} \rightarrow 0$, and identification of $\hat H_{0}(y_t)$ as the dashed curve in the right panel that describes inferred data hard components. 

\subsection{$\bf \eta$ density TCM with (almost) no assumptions}

This subsection responds to Ref.~\cite{bylinkin2014nucphb} which examines the role of quarks in nuclear collisions and wherein the thermal (exponential) element of the Bylinkin model is ``associated with the baryon valence quarks and a quark-gluon cloud coupled to the valence quarks. Those partons preexist long time before the interaction and could be considerer as being a thermalized statistical ensemble.'' The emphasis in  Ref.~\cite{bylinkin2014nucphb} is on $\eta$ distributions. The text below briefly summarizes the process of inferring TCM model functions from 200 GeV \pp\ $\eta$ density distributions without {\em a priori} assumptions as in Ref.~\cite{ppquad}.

A difference procedure as in Eq.~(\ref{etahcx}) can be used to derive a TCM for particle densities on pseudorapidity $\eta$. Given a 2D particle density on $(p_t,\eta)$ for some event class corresponding to particle multiplicity \nch\ and denoted by $\rho_0(p_t,\eta;n_{ch})$, a 1D density on \pt\ $\rho_0(p_t;n_{ch},\Delta \eta)$ must be obtained by {\em averaging} over some $\eta$ acceptance $\Delta \eta$ to obtain a \pt\ spectrum TCM described per Eq.~(\ref{tcmspec}) by
\bea \label{ytspeceq}
\bar \rho_0(p_t;n_{ch},\Delta \eta) &\approx& \bar \rho_{s}(n_{ch},\Delta \eta) \hat S_0(p_t)  
\\ \nonumber
&+& \bar \rho_{h}(n_{ch},\Delta \eta) \hat H_0(p_t,n_s),
\eea
where $\bar \rho_{x}(n_{ch},\Delta \eta) \equiv n_x / \Delta \eta$ with $x = s$, $h$ and bars indicate averages over $\Delta \eta$. That model, consistent with results from Ref.~\cite{ppprd}, describes corrected and extrapolated \pt\ spectra. 
For a \pt-integral study of densities on \etaa\ one may integrate $ \rho_0(p_t,\eta;n_{ch})$ over $p_t \in [0,\infty]$ to obtain
\bea \label{rho0eta}
\rho_0(\eta;n_{ch}) &\approx&   \rho_{s0}(n_{ch}) S_0(\eta)  
%\\ \nonumber
+  \rho_{h0}(n_{ch})  H_0(\eta),
\eea 
where $S_0(\eta)$ and $H_0(\eta)$ are TCM model functions to be determined as described below. 
Averaging Eq.~(\ref{rho0eta}) over some acceptance $\Delta \eta$ symmetric about $\eta = 0$ gives
\bea \label{etaav}
\bar \rho_0(n_{ch},\Delta \eta) &=&  \rho_{s0}(n_{ch}) \bar S_0(\Delta \eta) 
+ \rho_{h0}(n_{ch}) \bar H_0(\Delta \eta)
 \nonumber \\
&\equiv& \bar  \rho_{s}(n_{ch},\Delta \eta)  + \bar \rho_h(n_{ch},\Delta \eta),
\eea
with mean values $\bar \rho_x(n_{ch},\Delta \eta) = \rho_{x0} \bar X_0(\Delta \eta)$. The second line of Eq.~(\ref{etaav}) is then consistent with integrating Eq.~(\ref{ytspeceq}) over \pt. The TCM of Eq.~(\ref{rho0eta}) can be rewritten 
\bea \label{rho0eta2}
\rho_0(\eta;n_{ch},\Delta \eta) &\approx&  \bar  \rho_{s}(n_{ch},\Delta \eta) \tilde S_0(\eta;\Delta \eta)  
\\ \nonumber
&+& \bar  \rho_{h}(n_{ch},\Delta \eta)  \tilde H_0(\eta;\Delta \eta),
\eea 
with $\bar \rho_{s}(n_{ch},\Delta \eta)$ and $\bar \rho_{h}(n_{ch},\Delta \eta)$ inferred from measured $n_{ch}$ based on an assumed value for parameter $\alpha$ as discussed in Sec.~\ref{subtract} above.  The $\tilde X_0(\eta;\Delta \eta)$ (denoted by a tilde) have {\em mean value} 1 over acceptance $\Delta \eta$ by definition.  
Averaging Eq.~(\ref{rho0eta2}) over $\eta$ acceptance $\Delta \eta$ should then be consistent with integrating Eq.~(\ref{ytspeceq}) over \pt.

\begin{figure}[h]
	\includegraphics[height=1.6in]{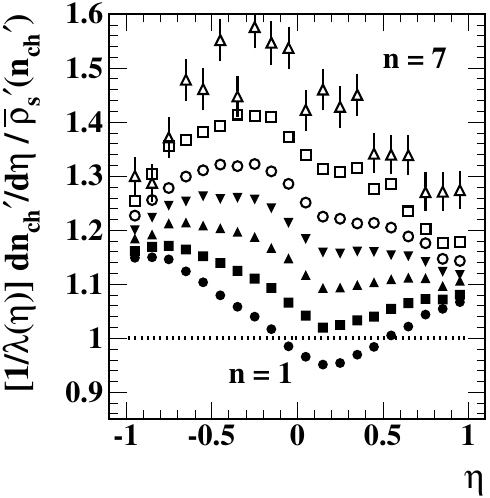}
	\includegraphics[height=1.6in]{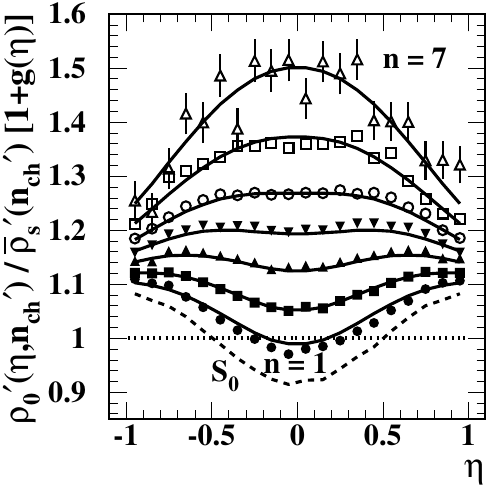}
\put(-129,85) {\bf (a)}
	\put(-18,85) {\bf (b)}\\
	\includegraphics[height=1.6in]{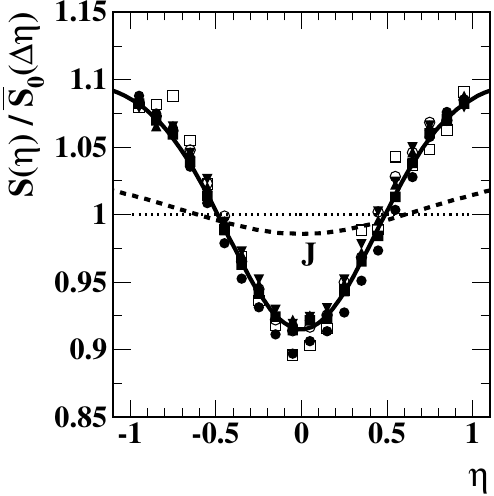}
	\includegraphics[height=1.6in]{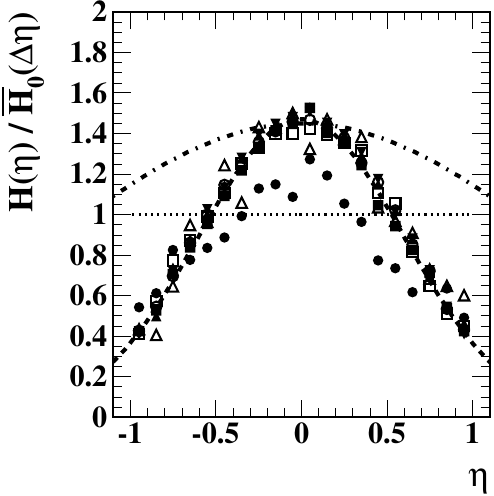}
	\put(-133,90) {\bf (c)}
	\put(-20,90) {\bf (d)}
	\caption{\label{etaquad}
(a) Uncorrected $\eta$ densities within $\Delta \eta = 2$ for seven multiplicity classes of 200 GeV \pp\ collisions~\cite{ppquad}.
(b)  Corrected $\eta$ densities within $\Delta \eta = 2$ for seven multiplicity classes. The dashed curve is normalized soft-component model $\tilde S_0(\eta) \equiv S_0(\eta)/ \bar S_0(\Delta \eta)$ from Eq.~(\ref{s00}). The solid curves are Eq.~(\ref{rho0eta2}) with TCM elements defined in Eqs.~(\ref{h00}) and (\ref{s00}).
(c) Data soft components inferred as in Eq.~(\ref{ss}). The solid curve is normalized soft-component model $\tilde S_0(\eta)$ as defined in Eq.~(\ref{s00}). The dashed curve is Eq.~(\ref{jacob}), the Jacobian from $y_z$ to $\eta$. 
(d) Data hard components inferred as in Eq.~(\ref{hc01}). The solid curve is normalized hard-component model $\tilde H_0(\eta)$ defined in Eq.~(\ref{h00}). The data for $n = 1$ (solid dots) are significantly low compared to the common trend. The dash-dotted curve is an estimate for the Bylinkin power-law model element reported in Ref.~\cite{bylinkin2014nucphb}, with $\sigma_{h} \approx 1.46$ for 200 GeV data.
	}  % ppcms111anoy, ay, dy, c2y
\end{figure}

Figure~\ref{etaquad} (a) shows uncorrected densities $\rho_0'(\eta;n_{ch})$ rescaled by particle density soft component $\bar \rho_s$ for seven event classes. The data have been efficiency corrected by factor $\lambda(\eta)$ but there is a substantial asymmetry on \etaa\ due to different readout chamber efficiencies in the two TPC end caps. Figure~\ref{etaquad} (b) shows data from panel (a) corrected by factor $1+g(\eta)$ that removes the asymmetry.

The full procedure for defining an $\eta$-density TCM is recursive: The formula immediately below is applied to obtain data hard components from which a hard-component model $\tilde H_0(\eta;\Delta \eta)$ is inferred. That model is then combined with data $\rho_0(\eta;n_{ch})$ to obtain soft components  from which soft-component model $\tilde S_0(\eta;\Delta \eta)$ is obtained.

To derive hard-component model function $\tilde H_0(\eta;\Delta \eta)$ the same model-independent subtraction technique as for \pt\ spectra in Sec.~\ref{subtract} is employed.  The results of  that first step are not shown here but may be found in Ref.~\cite{ppquad}. Data differences are determined as
\bea \label{etahc}
\tilde H_{0n}(\eta;\Delta \eta) &\equiv& \frac{\rho_0(\eta)_{n} /   \bar \rho_{s,n} - \rho_0(\eta)_{n-1} /  \bar \rho_{s,n-1}} {\bar \rho_{h,n} / \bar \rho_{s,n} - \bar \rho_{h,n-1} /   \bar \rho_{s,n-1}},
\eea
where index $n \in [1,7]$ represents seven multiplicity classes. As noted, the difference in the numerator cancels common term $\tilde S_0(\eta;\Delta \eta)$ (defined below). For $n = 1$ it is assumed that $\rho_0(\eta)_{n-1}/ \bar \rho_{s,n-1} \rightarrow \tilde S_0(\eta;\Delta \eta)$, and $\bar \rho_{h,n} / \bar \rho_{s,n} - \bar \rho_{h,n-1} / \bar  \rho_{s,n-1} \rightarrow \bar \rho_{h,1} / \bar \rho_{s,1}$ (i.e.\  extrapolation to $n = 0$ is assumed to be pure soft component and $\bar \rho_{h,0} = 0$). 
Given Eq.~(\ref{rho0eta2}) as TCM that expression represents a common {\em data} hard component $H_0(\eta) / \bar H_0(\Delta \eta)$. The inferred model function for $\Delta \eta = 2$ is 
\bea \label{h00}
\tilde H_0(\eta;\Delta \eta) \equiv \frac{H_0(\eta)}{\bar H_0(\Delta \eta)} &=& 1.47 \exp[-(\eta / 0.6)^2/2],
\eea
 where the ``normalized'' model function is denoted by a tilde. The data hard-component form is, within statistical errors, approximately independent of \nch\ over an interval  implying a {\em 100-fold increase} in dijet production, and suggests that most of the hard-component yield  (MB dijet fragments) falls within the acceptance $\Delta \eta = 2$.

%%%%%%%%%%
%%%%%%%%%%%%

Figure~\ref{etaquad} (c) shows {\em data} soft-component estimator
\bea \label{ss}
\tilde S_0(\eta;\Delta \eta) &\equiv& \rho_0(\eta;n_{ch}) /  \bar  \rho_{s} - (\bar \rho_{h} / \bar \rho_{s}) \tilde H_0(\eta;\Delta \eta),
\eea
with $\tilde H_0(\eta;\Delta \eta)$ as defined in Eq.~(\ref{h00}).  The inferred soft-component model for $\Delta \eta = 2$ (solid curve) is defined by
\bea \label{s00}
\tilde S_0(\eta;\Delta \eta)\hspace{-.02in} \equiv\hspace{-.02in} \frac{S_0(\eta)}{\bar S_0(\Delta \eta)}\hspace{-.05in} &=&\hspace{-.05in} 1.09 - 0.18 \exp[-(\eta / 0.44)^2/2].~~~~~
\eea
The form of the soft component also appears to be invariant over a large event \nch\ interval. Small data deviations from the model are consistent with statistical errors. A minimum at $\eta = 0$ is expected given  the Jacobian for $\eta \leftrightarrow y_z$, where an approximately uniform distribution on $y_z$ is expected within a limited $\Delta y_z$ acceptance. The observed excursion appears too large as discussed below.

Figure~\ref{etaquad} (d) shows data hard components estimated with an alternative method assuming $\tilde S_0(\eta)$ is known
\bea \label{hc01}
\tilde H_0(\eta;\Delta \eta) &=&  \frac{\rho_0(\eta;n_{ch}) /   \bar \rho_{s} - \tilde S_0(\eta;\Delta \eta)} {\bar \rho_{h} / \bar \rho_{s}},~~~
\eea
which substantially reduces statistical noise in  the differences. The dashed curve is the hard-component model $\tilde H_0(\eta;\Delta \eta)$ defined in Eq.~(\ref{h00}) confirming TCM self-consistency. The points for peripheral $n=1$ (solid dots) are low compared to the general trend, which may indicate that the hard component for $n=1$ extends low enough on $y_t$ to be reduced by low-\pt\ tracking inefficiency. The results in Fig.~\ref{etaquad} (c,d) indicate that soft and hard densities $\bar \rho_{x}(n_{ch};\Delta \eta)$ depend substantially on detector $\eta$ acceptance $\Delta \eta$. The dash-dotted curve is an estimate for the Bylinkin power-law model element reported in Ref.~\cite{bylinkin2014nucphb}, with $\sigma_{h} \approx 1.46$ for 200 GeV data. The amplitude is adjusted for tangency to the data at maximum.

The Jacobian relating densities on $\eta$ and $y_z$ is
\bea \label{jacob}
J(\eta) &=& \frac{\cosh(\eta)}{\sqrt{a^2 + \cosh^2(\eta)}}
\eea
with $a = m_{0i} / p_t$. For pions \mpt\ $\approx 0.45$ GeV/c and $m_{0} \approx 0.14$ GeV/$c^2$ so $a \approx 1/3$. Assuming uniform (boost invariant) density on rapidity $y_z$ within acceptance $\Delta \eta = 2$ the dashed curve on $\eta$ in panel (c) is then predicted. It is thus unlikely that the data trend in panel (c) is simply a consequence of the Jacobian. A possible source of the large excursion, transport of hadrons from soft to hard component that conserves species identities and follows statistical-model predictions, is described in Ref.~\cite{transport}, and see the discussion in Sec.~\ref{omega}.

%%%%%%%%%%%%
\section{Spectrum Models and predictivity} \label{predict}

This section demonstrates that the TCM describes high-mass hadron spectra in \pp\ (and other) collisions accurately and reveals new aspects of hadron production. The TCM for higher-mass hadrons is {\em predicted} based on trends determined from lower-mass hadrons. The Bylinkin model in contrast has only one effective element (power law) to describe such spectra, and the model parameters are not predicted or interpretable {\em within the context of the Bylinkin model itself}. Parameter trends {\em may} be interpreted in the larger context of the TCM.

\subsection{13 TeV p-p collisions and multistrange hadrons}

\begin{figure}[h]
	\includegraphics[width=1.65in]{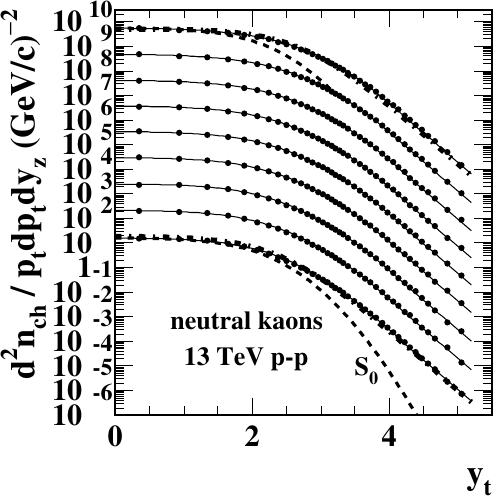}
	\includegraphics[width=1.65in]{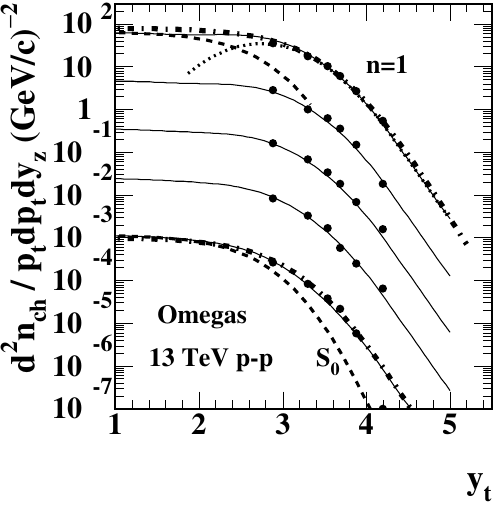}
	\caption{\label{piddata}
		\pt\ spectrum data (points) for strange hadrons:
		Left: Neutral kaons from Ref.~\cite{alicepppid}.
		Right: Omegas from Ref.~\cite{alippss}.
		Solid curves represent the PID spectrum TCM. 
Dashed curves are TCM soft components in the form $z_{si}(n_s) \bar \rho_s \hat S_0(y_t)$.  
The dotted curve at right is a TCM hard component in the form $z_{hi}(n_s) \bar \rho_h \hat H_0(p_t)$.
The dash-dotted curves are Bylinkin power-law model elements with three parameters adjusted to best describe data for each of two event classes in either panel. 
	}  % alippss1aay, 1adx
\end{figure}

Figure~\ref{piddata} shows spectrum data (points) for neutral kaons (left) and Omegas (right) from 13 TeV \pp\ collisions reported in Ref.~\cite{alippss}, here as densities on \pt\ vs transverse rapidity \yt\ with pion mass assumed. The spectra are rescaled by powers of 10 according to $10^{10-n}$ where $n \in [1,9]$ is the centrality class index and $n = 1$ is most central. Solid curves are full TCM parametrizations~\cite{ppsss}.  Dashed curves are TCM soft components in the form $ z_{si }\bar \rho_s \hat S_{0i}(m_t)$. Soft-component parameters for all hadrons except pions are $T \approx 200$ MeV and $n \approx 14$. Variation of $\hat S_{0i}(m_t)$ with hadron species is entirely due to mass dependence of \mt\ -- see Fig.~\ref{shape} (right). Data$-$soft-component differences represent minimum-bias jet contributions modeled by hard component $ z_{hi} \bar \rho_h \hat H_{0i}(p_t)$ with $\hat H_{0i}(y_t)$ (as density on \yt) parameters $\bar y_t = 2.7 $, $\sigma_{y_t}= 0.58 $ and $q= 3.7$ for kaons and $\bar y_t = 3.0 + 0.55 x(n_s)$, $\sigma_{y_t}= 0.50$ and $q= 4.6$ for Omegas. Factors $z_{xi}$ are fractions ($\leq 1$) of soft and hard densities $\bar \rho_s$ and $\bar \rho_h$ for unidentified hadrons determined previously~\cite{ppbpid}, and $x(n_s) = \bar \rho_h / \bar \rho_s$.

%%%%%%%%%%
%%%%%%%%%%%%

Also shown are two dash-dotted curves in each panel based on the Bylinkin model power-law element as it appears in Eq.~(\ref{byeq}). The parameters $T_h$ and $N$ and constant multiplier $A_h$ are freely adjusted to best describe the data. For central ($n = 1$) and peripheral (n = 5 or 9) data respectively the $T_h$ parameter values are 0.66 and 0.48 GeV/c for kaons and 1.44 and 1.04 GeV/c for Omegas. Exponent $N$ values are 2.8 for kaons and 3.8 for Omegas with no significant variation with event class. While that three-parameter model does describe spectra for higher-mass hadrons reasonably well, interpretation of parameter values is problematic. 
Note that the Tsallis model assumes $T$ is a thermodynamic temperature. Within \pp\ collisions do different hadron species in different event classes experience different temperatures?

In contrast, the TCM is simply interpreted in terms of standard QCD processes and measured jet properties. Figure~\ref{shape} (right) demonstrates that the apparent shoulder of soft-component $\hat S_0(m_t)$ extends to higher \pt\ with increasing hadron mass, accounting for some of the $T_h$ variation in the Bylinkin power-law model (i.e.\ from kaons to Omegas). Variation of $N$ corresponds to variation of fragmentation functions with hadron species as noted in Ref.~\cite{ppbpid}. In addition, jet production increases quadratically with increasing event \nch\ which further drives the {\em apparent} data shoulder to higher \pt\ because of soft- and hard-component interaction, thus explaining variation with event class for $T_h$ but {\em lack thereof for $N$}.

\subsection{Omega hard/soft ratios and hadron mass} \label{omega}

Systematic TCM analysis of a variety of collision systems and hadron species leads to new insights into collision dynamics. In particular, analysis of PID spectra including multistrange hadrons reveals close correlations between soft and hard components that relate to the statistical model of hadron production. Those result demonstrate that soft and hard production are quantitatively described {\em and predicted} by the TCM, and the relation between the two is also established. Such results demonstrate what is missed by single-element or monolithic spectrum models whose parameters are uninterpretable.

In Refs.~\cite{ppbpid,pidpart1,pidpart2,pppid}  soft and hard fractions $z_{si}(n_s)$ and $z_{hi}(n_s)$ (where $n_s$ is the soft particle multiplicity used as an event index and $i$ denotes hadron species) noted in the previous subsection are inferred from 13 TeV \pp\ and 5 TeV \ppb\ PID spectra. From the algebraic forms inferred from data, ratios $\tilde z_i(n_s) \equiv z_{hi}(n_s)/z_{si}(n_s)$ are observed to be approximately independent of $n_s$ and {\em linearly proportional to hadron mass} {\em irrespective of baryon identity or strangeness}.  In Ref.~\cite{transport} that information is combined to determine that density ratios $\bar \rho_{si}/\bar \rho_{0i}$ and $\bar \rho_{hi}/\bar \rho_{0i}$ are simply related. The result has major implications for the spectra of more-massive hadrons.

\begin{figure}[h]
	\includegraphics[width=1.65in]{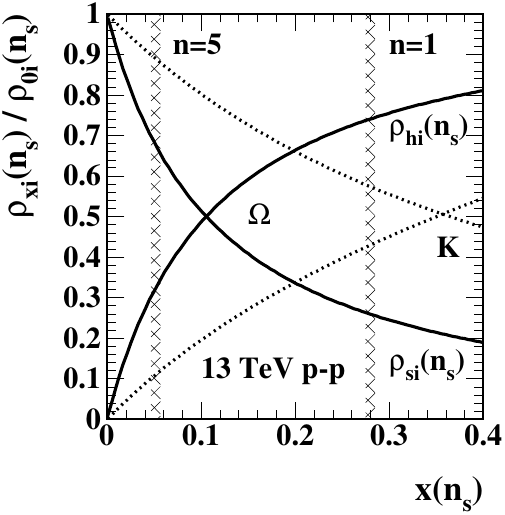}
	\includegraphics[width=1.65in]{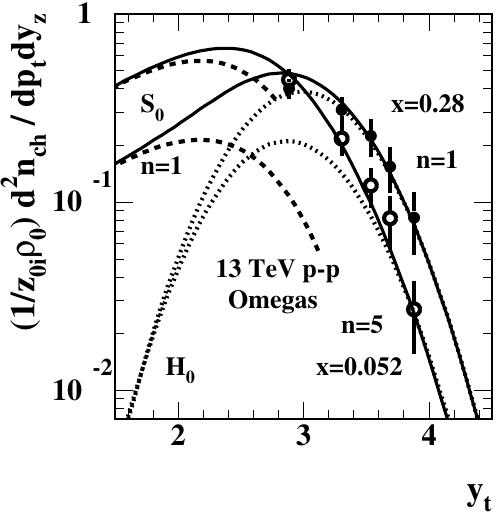}
	\caption{\label{omegarats}
		Left: Hard- and soft-component yield ratios $\bar \rho_{si} / \bar \rho_{0i}$ and $\bar \rho_{hi} / \bar \rho_{0i}$ for 13 TeV \pp\ collisions from Ref.~\cite{transport} providing a {\em prediction} (curves) for Omega baryons based on extrapolation of a mass trend for lower-mass hadrons. The hatched bands correspond to hard/soft ratio $x(n_s)$ values for Omega spectra from \pp\ event classes 1 (high \nch) and 5 (low \nch). The curves are labeled with numerators on the $y$ axis.
		Right: PID spectra (points) for Omega baryons for event classes 1 and 5 from 13 TeV \pp\ collisions~\cite{ppsss} plotted per the y-axis label vs pion transverse rapidity \yt. The solid curves are TCM spectra  rescaled per the $y$-axis label. The dashed and dotted curves are corresponding soft and hard TCM model functions.
	} %  alipid5ex, alippss5a
\end{figure}

Figure~\ref{omegarats} (left) shows results from Ref.~\cite{transport}.  The curves correspond to {\em predicted} ratios $\bar \rho_{si}/\bar \rho_{0i}$ and $\bar \rho_{hi}/\bar \rho_{0i}$ vs hard/soft ratio $x(n_s) = \bar \rho_h(n_s)/\bar \rho_s(n_s) \propto \bar \rho_s$ for kaons (dotted) and Omega baryons (solid) from 13 TeV \pp\ collisions. The single control parameter for those trends is ratio $\tilde z_i \propto m_{0i}$. This plot illustrates how low-\pt\ behavior [$\bar \rho_{si}(n_s) = z_{si}(n_s)\bar \rho_s$] predicts high-\pt\ behavior [$\bar \rho_{hi}(n_s) =z_{hi}(n_s)\bar \rho_h$], for example enabling correction of proton inefficiencies in Refs.~\cite{pidpart1,pppid}. It also illustrates the very strong dependence of hard/soft ratios on hadron mass. Hadron production mechanisms for jet and nonjet components of the final state are closely correlated so as to preserve the {\em total} hadron number or fraction for each species consistent with statistical-model predictions~\cite{statmodel}. 

The hatched bands correspond to values of hard/soft ratio $x(n_s)$ for 13 TeV \pp\ event classes 1  (0.28, high \nch) and 5 (0.052, low \nch) for Omega spectra. For event class 5 more  than two-thirds of Omegas are nonjet (projectile proton fragments) while for event class 1 {\em three quarters of Omegas are jet fragments}, thus representing a dramatic change in hadron production while still maintaining approximately the same {\em total} Omega fraction $z_{0i}^* \approx 0.0004$. The actual fractions $z_{0i}(n_s)$ used are consistent with \pp\ Omega fraction data in Fig.~6 (right) of Ref.~\cite{ppsss}.

%%%%%%%%%%
%%%%%%%%%%%%

Figure~\ref{omegarats} (right) shows Omega spectra (points) for \pp\ event classes 1 and 5 in the form $(1/z_{0i} \bar \rho_0) d^2n_{chi} / dp_t dy_z$ without the usual additional factor $1/p_t$, better indicating what fraction of particles lies within a given \pt\ interval. In that format the TCM spectra each integrate on \pt\ to 1. The error bars are published total systematic uncertainties. The solid curves are TCM spectra appearing in Fig.~\ref{piddata} (right) rescaled in the same way. 
The dashed curves are soft components with $T = 200$ MeV and $n = 14$, and the dotted curves are corresponding TCM hard components (jet fragment distributions) with $\bar y_t = 3.03,~3.15$, $\sigma_{y_t} = 0.5$ and $q = 4.6$ ($\sim 1/p_t^{6.6}$). 

As in the left panel the integrated hard component is three times the integrated soft component for event class 1, and data points lie essentially entirely on the hard component. Thus, jets dominate Omega production from \pp\ collisions for event class $n = 1$. That is why in Fig.~\ref{piddata} (right) the monolithic $q$-Gaussian for $n = 1$ (dash-dotted) lies well above the TCM (solid) at low \yt\ whereas it lies significantly below the TCM for $n =  5$.
The TCM for Omega spectra is based on {\em predictions} from Ref.~\cite{transport} where it is determined that fractional coefficient $z_{0i}$ controls integrated yields from {\em entire} hadron spectra, {\em including jet contributions} at high \pt, while the TCM accurately describes dramatic changes in {\em differential} spectrum structure in terms of two fragmentation processes. 

\subsection{Ensemble-mean $\bf \bar p_t$ systematics}

This subsection responds to  ensemble \mpt\ or \mmpt\ results reported in Ref.~\cite{bylinkinprd}, in particular regarding its Fig.~6, and in Ref.~\cite{bylinkin2016b}. The Bylinkin \mpt\ trends vs $\bar \rho_0$ and $\sqrt{s}$ are discussed in Sec.~\ref{mpttrends} above. Those trends are then combined to produce curves for \mpt\ vs \nch\ at several collision energies in Fig.~6 of Ref.~\cite{bylinkinprd}. The paper concludes that ``Accounting for mini-jet contribution the \mpt\ should increase with \nch\ since another way to enlarge multiplicity is to produce mini-jets with a larger $E_t$ (and thus a larger multiplicity in jet-fragmentation).'' Although the composite Bylinkin model curves are reasonably close to data the individual elements of the Bylinkin model are problematic as illustrated in Figs.~\ref{charge} and \ref{mptby} above. The former demonstrates that Bylinkin charge densities exhibit strong cross talk between soft and hard elements vs \nch, and the energy dependence deviates strongly from measured total charge densities on collision energy. The latter demonstrates that power-law \mmpt\ is intermediate between TCM soft and hard components because of strong overlap between two Bylinkin elements.

Reference~\cite{bylinkin2016b} focuses on \pt\ spectra for baryons where it concludes that the exponential (soft) Bylinkin element is not required by data, so only the Bylinkin power-law function is implemented. Three models are applied to data: an isolated exponential (QGSM), a Tsallis function and the Bylinkin power-law element. It is asserted that ``these three approaches have the exponential-like behavior in the low-\pt\ region...'' But the power-law element is a Gaussian at low \pt. Apparently any model may be applied to determine \mmpt\ for baryons ``...since high-\pt\ particles...do not give a big contribution to the mean value.'' 
That remark conflicts with the conclusion ``The enhancement of power-low [sic] contributions into the spectra at high \pt's [sic] causes the change of low \pt\ exponential slopes, so that [\mpt] are growing with energy.''   Those statements conflict with TCM results as shown below.

In Fig.~\ref{mptby} (right) above, the Bylinkin $\langle p_t \rangle_{hb} \propto (\sqrt{s})^{0.07}$ trend~\cite{bylinkinprd} is intermediate between TCM hard and soft components. As noted, the power-law model confuses two production mechanisms. The TCM hard-component trend (upper solid) is $\langle p_t \rangle_{h} \propto (\sqrt{s})^{0.05}$. In Ref.~\cite{bylinkin2016b} the trend for baryons in its Fig.~2 is given as $\langle p_t \rangle_{hb} \propto s^{0.055}$ (confirmed from that figure in the present study, note $s^x$ vs $(\sqrt{s})^x$) which is said to be {\em the same as the trend for pions} in Ref.~\cite{bylinkinprd}, presenting an unresolved contradiction.

Reference~\cite{bylinkin2016b} also considers  the hadron mass-dependence of \mmpt\ trends. Based on its Fig.~3 it concludes that  ``These [\mpt] are also growing linearly [with mass] if [one is analyzing different baryon spectra based on masses].'' Note that statements in Ref.~\cite{bylinkin2016b} are based on proton spectra for $p_t \in [0.35,2]$ GeV/c, a restricted interval. The simple statement conceals more-detailed relationships revealed by TCM analysis. An example is provided by study of \mmpt\ for multistrange hadrons below.

The TCM form for the conventional \mpt\ or \mmpt\ ratio
\bea \label{ppmpttcm}
\bar p_t' \equiv \frac{\bar P_t'} {n_{ch}'} &\approx & \frac{\bar p_{ts} + x(n_s) \bar p_{th}(n_s)}{\xi + x(n_s)},
\eea
where $x(n_s) \equiv n_{h} / n_{s} \approx  \alpha \bar \rho_s$ is the ratio of hard-component to soft-component yields~\cite{ppprd} and $\xi \leq 1$ accounts for a low-\pt\ spectrum acceptance cutoff, conflates two simple TCM trends.  Alternatively, the ratio
\bea \label{niceeq}
\frac{n_{ch}'}{n_s} \bar p_t'   \approx \frac{ \bar P_t}{n_s} &= & \bar p_{ts}(\sqrt{s}) + x(n_s) \bar p_{th}(n_s,\sqrt{s})
%\\ \nonumber
%&\approx& \bar p_{ts} + \alpha(\sqrt{s})\, \bar \rho_s \, \bar p_{th}(n_s,\sqrt{s}),
\eea
where $n_{ch}' = \xi n_s + n_h$,
provides a simple and convenient basis for testing the TCM hypothesis precisely. Note that $\bar p_{ts}$ and $\bar p_{th}$ depend on collision energy per Fig.~\ref{mptby} (right). That dependence is left implicit in what follows.

Equation~(\ref{niceeq}) describes TCM \mmpt\ trends for unidentified hadrons. For identified hadrons of species $i$ from \pa\ collisions ($\nu \geq 1$) the corresponding description is
\bea \label{pampttcmpid}
\frac{\bar P_{ti}}{n_{si}} &=& \frac{n_{chi}}{n_{si}}\bar p_{ti} \approx \bar p_{tsi} + \tilde z_i(n_s) x(n_s)\nu(n_s) \, \bar p_{thi}(n_s).~~~
\eea
Factors $\tilde z_i(n_s) x(n_s)$ have been defined above , $\nu(n_s) \equiv 2 N_{bin} / N_{part}$ (binary collisions per participant pair) is a geometry parameter for \pa\ and \aa\ collisions, and soft- and hard-component mean values $\bar p_{tsi}$ and $\bar p_{thi}(n_s)$ are obtained from TCM models $\hat S_{0i}(y_t)$ and $\hat H_{0i}(y_t,n_s)$, where $\hat H_{0i}(y_t,n_s)$ may be slowly varying with $n_s$~\cite{pidpart2,pppid}.

\begin{figure}[h]
	\includegraphics[width=3.3in]{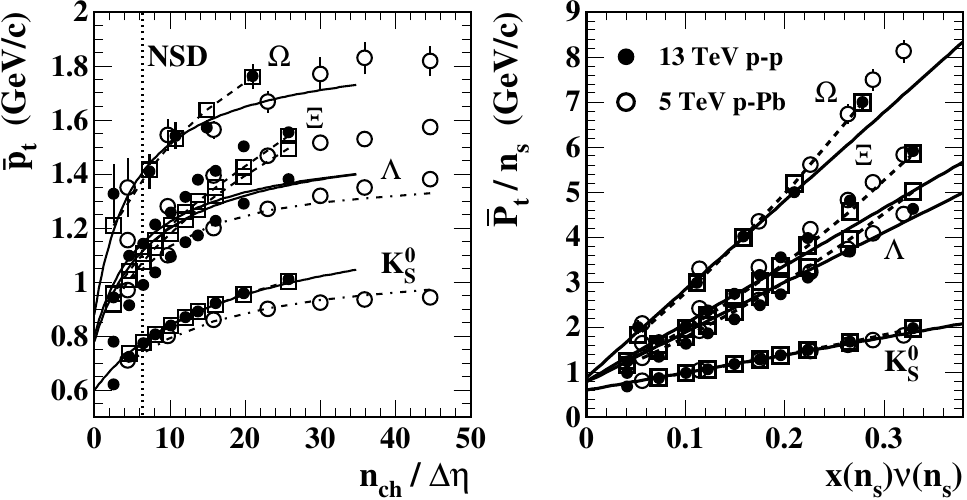}
	\caption{\label{mptss}
		Left: Ensemble-mean \mmpt\ data for 13 TeV \pp\ collisions from Fig.~5 of Ref.~\cite{alippss} (solid dots) and for 5 TeV \ppb\ collisions (open circles) from Ref.~\cite{aliceppbpid}. Open squares (connected by dashed curves) are \pp\ \mmpt\ values inferred from TCM spectra in Ref.~\cite{ppsss}. 
%		Open circles are 5 TeV \ppb\ data. 
		Dash-dotted curves are \ppb\ \mmpt\ TCM trends from Refs.~\cite{ppbpid,pidpart2}. 
		Right: Data and curves from (a) transformed to the TCM format of Eq.~(\ref{pampttcmpid}).
	}  % alippss2axx
\end{figure}

Figure~\ref{mptss} (left) shows \mmpt\ data for neutral kaons, Lambdas, Cascades and Omegas in a conventional \mmpt\ format equivalent for PID data to Eq.~(\ref{ppmpttcm}). The solid symbols are for 13 TeV \pp\ collisions, the open symbols for 5 TeV \ppb\ collisions. The line marked NSD is the NSD $\bar \rho_0 = n_{ch} / \Delta \eta$ value for 13 TeV \pp\ collisions.

%%%%%%%%%%
%%%%%%%%%%%%

Figure~\ref{mptss} (right) shows the same data and curves transformed to the format of Eq.~(\ref{pampttcmpid}) by multiplication with factor $1 + \tilde z_i(n_s) x(n_s) \nu(n_s)$ ($\nu \rightarrow 1$ for \pp\ data and $x(n_s) = \alpha \bar \rho_s$). In the format of Eq.~(\ref{pampttcmpid}) \pp\ and \ppb\ data are directly comparable, and the close correspondence between the two systems (note open circles vs solid dots) is evident, whereas that is not the case at left. The solid lines are a TCM with fixed hard components. The dashed curves and open boxes include \pp\ hard components $\bar p_{thi}(n_s)$ that vary systematically with event class per $x(n_s)$ to accommodate data~\cite{ppsss}.  

The slopes in the right panel are determined by product $ \tilde z_i(n_s)\bar p_{thi}(n_s)$ per Eq.~(\ref{pampttcmpid}). While $\bar p_{thi}(n_s)$ may vary by a few tens of percent from one hadron species to another (and {\em not} monotonically with mass) it is hard/soft fraction ratio $\tilde z_i(n_s)$ that varies directly proportional to hadron mass $m_{0i}$ (cf.\ Sec.~\ref{omega}), leading to  {\em factor 30} increase in the product (slope) from pions to Omegas. Thus, it is  mass dependence of jet {\em number}, not hard-component {\em shape}, that dominates \mmpt\ variation with hadron species. The \mmpt\ mass trend reported in Fig.~3 of Ref.~\cite{bylinkin2016b} may be compared with values at the dotted line of Fig.~\ref{mptss} (left). In particular, the values for pions and kaons in Ref.~\cite{bylinkin2016b}  seem to have large systematic errors.

To summarize from Figure~\ref{mptss} (right), all hadron species require distinct soft and hard spectrum components inferred from data as in Fig.~\ref{none}. The Bylinkin exponential mimics the former to a greater or lesser extent depending on hadron mass and jet/nonjet ratio. \mmpt\ soft component $\bar p_{tsi}$ is mass dependent because of required evolution of the $\hat S_0(m_t)$ model function with hadron mass as exhibited in Fig.~\ref{shape} (right). Higher-mass hadrons, i.e.\ baryons, require distinct model representation and make large contributions to \mmpt\ as demonstrated above.

%%%%%%%%%%%%%%%%%%
\section{Thermalization mechanisms} \label{thermal}

Inference of ``collective phenomena'' in small collision systems since 2010  has generated interest in possible mechanisms to achieve thermalization in such systems. The Bylinkin spectrum model and interpretation of its exponential element as an indicator for thermalization plus a proposal of quantum entanglement as a mechanism for thermalization have drawn significant attention.

\subsection{Thermalization within small collision systems} \label{small}

As noted, a chronic problem for the nuclear physics community over the past decade has been appearance in small (\pp, \pa) collision systems at the LHC of ``flow-like'' phenomena interpreted as such for \aa\ collisions~\cite{hydro}. Either those phenomena are not  manifestations of flows, which jeopardizes QGP formation claims for \aa\ collisions, or a dense medium {\em is} generated even in smaller systems whose gradients then drive the claimed flows.
Resolution of that dilemma motivates introduction of {\em ad hoc} mechanisms that either rely on final-state particle rescattering in a low-density environment to produce a thermalized high-density medium or invoke some more-exotic strategy. The narrative outlined below appears to fall into the latter category, relying on the Bylinkin-model exponential element to signal thermalization. 

In Ref.~\cite{bylinkintherm} a novel thermalization mechanism is based on the Bylinkin spectrum model and its response to various experimental situations in terms of the perceived presence or absence of its exponential element. The argument proceeds with ``The [Bylinkin model] thermal component  is due to the effective event horizon introduced by the confining string.... The slope [$T_{th}$] of the soft component of the hadron spectrum ... is determined by the saturation momentum that drives the deceleration in the color field....'' Those statements are associated with an Unruh effect conjectured to arise from acceleration in a vacuum, possibly related to Hawking radiation from black holes. Thermal emission of hadrons is linked to deceleration by color fields (strings) between partons.

The argument continues with ``The hard component is well understood as resulting from the high momentum transfer scattering of quarks and gluons, and their subsequent fragmentation. The `soft' one is ubiquitous in high energy collisions and has the appearance of the thermal spectrum -- but its origin remains mysterious to this day. ...while in nuclear collisions one may expect thermalization to take place, it is hard to believe that thermalization can occur in ... DIS or \ee\ annihilation. ... The universal thermal character of hadron [\pt] spectra and abundances in all high energy processes...begs for a theoretical explanation.'' 
The paper concludes with ``We hope  that our analysis sheds some new light on the origin of the thermal component in hadron production. ... the soft hadron production [Bylinkin exponential] is a consequence of the quantum evaporation from the event horizon formed by deceleration in longitudinal color fields.''

That narrative is based on several assumptions, (a) that the Bylinkin-model exponential term is demonstrably required by spectrum data, (b) that it actually represents hadron production from a thermalized emitter and (c) that there are no alternative interpretations of spectrum data available, thus requiring a novel explanation. 

Assumption (a) fails as demonstrated in previous sections: see Figs.~\ref{charge} and \ref{mptby} for Bylinkin model parameters that deviate strongly from trends required by data and other figures where  the exponential element is excluded. Assumption (b) fails as demonstrated for example by Fig.~\ref{eedata} (left) where hadron emission from \ee\ collisions (points) exhibits the same soft-component $\hat S_0(m_t)$  (solid curve) as that from \pp\ and \pa\ collisions that strongly diverges from a pure exponential (dotted curve). That difference indicates that the emitting system is {\em not} thermalized.  $\hat S_0(m_t)$ exponent $1/n > 0$ signals that the emitting system (parton splitting cascade) includes a noise source in the form of {\em Gribov diffusion}~\cite{gribov} equivalent to a random walk in transverse phase space of successive parton generations, deeper with increasing energy. Diffusion in configuration space leads to increasing inelastic cross sections~\cite{gribov3}. Diffusion in momentum space leads to increase of $\hat S_0(m_t)$ parameter $1/n$ in Fig.~\ref{power} (left)~\cite{gribov2}.
 
Assumption (c) collides with decades of work to understand, theoretically and experimentally, the role of QCD in the context of high-energy nuclear collisions. The theoretical basis for the TCM was already enunciated in Ref.~\cite{pancheri} (1985): The low-\pt\ part or soft component, what the Bylinkin exponential element is nominally intended to describe, ``...does not show any multiplicity dependence....'' (referring to $P_0(p_t)$ in Eq.~(\ref{lia})), consistent with TCM $\hat S_0(m_t)$ but not with the strongly varying $T_{th}$ of the Bylinkin exponential element.  Referring to UA1 measurements, that detector ``measures not only the hard parton scattering process but also {\em the debris which result from the breaking up of the [projectile] proton when a hard parton is emitted} [emphasis added]~\cite{pancheri}.'' The splitting cascade resulting in ``debris'' from projectile nucleon dissociation is the same process evident in Fig.~\ref{eedata} below.

Aside from the issues raised above there is one of consistency in assigning the Bylinkin model exponential element to indicate system thermalization via various means  according to various conditions. In Ref.~\cite{bylinkin2014nucphb} it is stated that ``Those partons preexist long time before the interaction and could be considered as being a thermalized statistical ensemble. The hadrons [what species?] from this source are distributed presumably according to the Boltzmann-like exponential...distribution....'' In Ref.~\cite{bylinkin2015} there appears ``The exponential term in this model is associated with thermalized production of hadrons by valence quarks and a quark-gluon cloud coupled to them.'' Such scenarios contrast with that proposed in Ref.~\cite{bylinkintherm} relying on ``the effective event horizon'' also noted  above.

Problems arise for a scenario of hadrons emerging from a thermal source when the Bylinkin exponential is limited to pions only: In Ref.~\cite{bylinkinanom}  ``...the spectra of charged kaons and protons...shows no room for the contribution of the exponential term'' with the comment ``...only pions require a sizable exponential term in their...spectra. This anomaly observed in pion production didn't find a consistent explanation yet.'' In Ref.~\cite{bylinkphen} ``...the $\pi$ production in $pp$ collisions is dominated by a release of quasi-thermalized particles, while the spectra of heavier [hadrons] are dominated by pQCD like production mechanisms, leaving a relatively small room for thermalized particle production.'' Apparently, pions arise from one production mechanism while more-massive hadrons arise from another. Referring to spectrum ratios in Ref.~\cite{bylinkphen} ``the observed significant difference in the exponential contributions to the hadron spectra implies the {\em difference in the hadron production mechanisms} rather than an artifact of  the fit procedure [emphasis added].''

\subsection{Thermalization and entanglement} \label{entangle}

Reference~\cite{dimaentangle} seeks to account for apparent thermalization in small collision systems (e.g.\ \pp\ collisions): ``...we test the hypothesis about the link between quantum entanglement and thermalization in high-energy collisions.'' Evidence for thermalization is inferred from a specific \pt\ spectrum model:  ``...we address the origin of the apparent thermalization in high-energy collisions that is {\em usually inferred from the presence of the exponential component} [of the Bylinkin model] in the [\pt] distributions of produced particles and the thermal abundances of the hadron yields [emphasis added].'' The latter phrase refers to observed agreement between data and statistical-model predictions for hadron species abundances~\cite{statmodel}.
Inferred thermalization has usually been attributed to particle multiple scattering: ``The emergence of the thermal features in a high energy proton-proton collision is surprising, as the number of secondary interactions in  this process is relatively low and does not favor thermalization through conventional final-state interaction mechanisms.'' An alternative mechanism is sought.

The authors invoke quantum entanglement: ``...we will investigate the possibility that the apparent thermalization in high energy collisions is achieved during the rapid `quench' induced by the  collision due to the high degree of entanglement inside the...colliding protons. ... Since a high-energy collision can be viewed as a rapid quench of the entangled partonic  state, it is thus possible that the effective temperature inferred ... can depend on the momentum transfer....''  Based on spectrum analysis with the Bylinkin model ``We confirm the relation between the effective temperature [$T_{th}$] and the hard scattering scale [$T_h$] observed at lower energies, and find that  it extends even to the Higgs boson production process.'' 

Within the context presented by Ref.~\cite{dimaentangle} the observed close coupling of  $T_{th}$ and $T_h$ is viewed as supporting a conjecture, but from a standpoint of general data modeling any close correspondence of parameters from distinct model elements signals redundancy. The two TCM model components, derived empirically from data trends, are completely independent in that regard. Such parameter independence indicates, as an experimental finding, that there are two distinct hadron production mechanisms.

\subsection{Entanglement and hadron-hadron collisions}

Reference~\cite{dimaentangle} analyzes data spectra from several collision systems with the Bylinkin model for requirement {\em or not} of the exponential model element: a hadron spectrum from nondiffractive 13 TeV \pp\ collisions, a dimuon spectrum from the process $\gamma$-$\gamma \rightarrow \mu^+$-$\mu^-$ for 13 TeV \pp\ collisions, and Higgs boson decays to $\gamma$-$\gamma$ and to four leptons.
Reference~\cite{dimaentangle} first applies the Bylinkin model to inelastic \pp\ collisions to confirm that exponential (``thermal'') and power-law (``hard'') model elements combined are able to describe spectrum data. That result is interpreted to confirm thermalization in such collisions.

\begin{figure}[h]
	\includegraphics[width=3.3in]{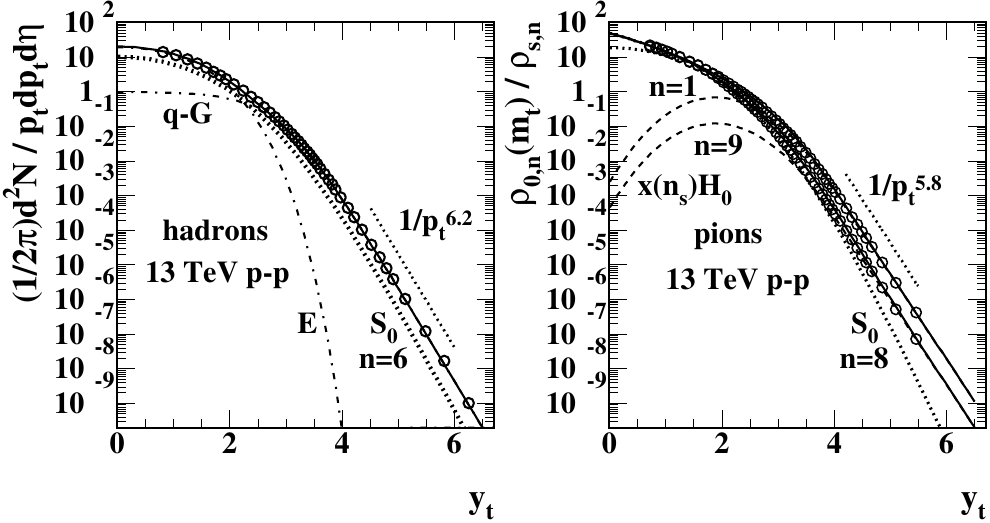}
	\caption{\label{ppcompare}
		Left: Minimum-bias unidentified-hadron spectrum from 13 TeV \pp\ collisions (points) and Bylinkin model elements exponential E and power-law ($q$-Gaussian) $q$-G (dash-dotted) with the sum (solid). A TCM soft-component model $\hat S_0(m_t)$ (dotted) with $n = 6$ (matching $N \approx 3$) is included for comparison. 
		Right: Identified-pion spectra from 13 TeV \pp\ collisions (points) for event classes $n = 1$ (high \nch) and $n  =9$ (low \nch). The spectra have been rescaled by TCM soft-component particle density $\bar \rho_s$. Model elements $\hat S_0(m_t)$ (dotted) and $x(n_s)\hat H_0(y_t)$ (dashed) are not fitted to individual spectra and constitute predictions. The full TCM (solid) includes a resonance contribution at low \pt\ (see Fig.~\ref{fac1}).
}  % acompare20
\end{figure}

Figure~\ref{ppcompare} (left) shows MB unidentified-hadron data from 13 TeV \pp\ collisions (points). The dash-dotted curves include a $q$-Gaussian labeled $q$-G and an exponential labeled E representing the two elements of the Bylinkin spectrum model as they appear in Fig.~2 of Ref.~\cite{dimaentangle}. The Bylinkin model Eq.~(\ref{byeq}) (solid, sum of two elements) describes the MB data reasonably well. The exponential element is determined by $T_e \leftrightarrow T_{th} = 0.17$ GeV which for this figure is adjusted to 0.145 GeV for better agreement with data, consistent with TCM analyses. Power-law parameter $T_h = 0.72$ GeV/c determines the shoulder position or width of the $q$-Gaussian. Parameter $N = 3.1$ determines the power-law exponent as $1/p_t^{2N} \rightarrow 1/p_t^{6.2}$ illustrated by the dotted line. The dotted curve is TCM $\hat S_0(m_t)$ with slope parameter $T = 0.145$ GeV and exponent $n = 6$ displaced down from data for visibility to demonstrate that a single function {\em can} describe those data. This comparison demonstrates that for MB data the Bylinkin model is sufficient, {\em but not necessary}. Monolithic model $\hat S_0(m_t)$ performs just as well.

%%%%%%%%%%
%%%%%%%%%%%%

Figure~\ref{ppcompare} (right) shows pion spectrum data (open circles) for low ($n  =9$) and high ($n = 1$) event multiplicity 13 TeV \pp\ collisions. The same data appear in Fig.~\ref{none} (left). The published spectrum data are presented in the form $\bar \rho_0(m_t) = d^2n_{ch}/p_t dp_t dy_z$ rescaled to $\bar \rho_0(m_t)/\bar \rho_s$ for pions, with $\bar \rho_s$ the root of $\bar \rho_0 = \bar \rho_s + \alpha \bar \rho_s^2$ and $\alpha \approx 0.017$ for 13 TeV \pp~\cite{alicetomspec}.  Since NSD  $\bar \rho_s \approx 5.8$ for 13 TeV is near factor $2\pi$ the unrescaled MB hadron spectrum at left is similar to $\hat S_0(m_t)$ (dashed) at right.
TCM parameters for $\hat S_0(m_t)$ are slope parameter $T = 0.145$ GeV and L\'evy exponent $n = 8.0$~\cite{alicetomspec} and for $\hat H_0(y_t)$ (defined as a density on \yt) are mode $\bar y_t = 2.45$, width above the mode $\sigma_{y_t} = 0.58$ and exponential parameter $q = 3.8$~\cite{pppid}. $\hat H_0(y_t)$ is here transformed to a density on \pt. The value for $q$ is thus related to the \pt\ power law as $2N \approx q +2$ because the Jacobian from density on \pt\ to density on \yt\ is $m_t p_t / y_t$ involving in effect a factor $p_t^2$ at higher \pt.

The spectrum rescaling $\bar \rho_0(m_t)/\bar \rho_s$ ensures that all spectra coincide at lower \pt\ where the hard/soft model-function ratio is $\ll 1$. Note that NSD $\bar \rho_s(\sqrt{s})$ is consistently defined from 13 TeV down to 64 GeV and is not a fit parameter, nor is parameter $\alpha(\sqrt{s})$. Likewise, exponent $n(\sqrt{s})$ is determined by the expression in Fig.~\ref{power} (left). $\hat H_0(y_t)$ parameters for each hadron species (here pions) are also determined by previous studies and are not available as fit parameters for individual spectra. 

The right panel demonstrates that variation of pion spectrum shape with increasing event multiplicity consists entirely of different admixtures of {\em two fixed functions} where the admixture (i.e.\ two fractions) is predicted based on the relation $\bar \rho_h \approx \alpha(\sqrt{s}) \bar \rho_s^2$ or $x(n_s) \equiv \bar \rho_h / \bar \rho_s \approx \alpha(\sqrt{s}) \bar \rho_s$. Those two functional forms, determined empirically from data~\cite{ppprd}, can be seen as ``eigenfunctions'' for \pp\ spectrum description: Any \pp\ spectrum for a given collision energy may be expressed as a linear combination of the two. In contrast,  parameters $T_{th}$, $T_h$ and $N$ of the two elements of the Bylinkin model must change substantially with changing collision conditions (e.g.\ \nch). Their values are then not interpretable and the Bylinkin model is excluded. The TCM is shown to be sufficient {\em and necessary} -- thus statistically equivalent to \pp\ data.

\subsection{Entanglement and exotic systems} \label{exotic}

Having concluded that Bylinkin-model fits to hadron spectra from MB 13 TeV \pp\ collisions demonstrate the existence of an exponential element, and therefore evidence for thermalization, Ref.~\cite{dimaentangle} considers a scenario where thermalization is not expected via entanglement --  a dimuon spectrum from the process \mbox{$\gamma$-$\gamma$} $\rightarrow \mu^+$-$\mu^-$. It is remarked that ``such diffractive events are expected to have a suppressed thermal (exponential) component. ... As the presence of the thermal component in this approach is the consequence of the entanglement, we expect it to be absent in diffractive events.'' It is notable  that events accepted require that {\em both protons remain intact}. Thus, the statement ``...these diffractive processes allow an analysis of the fragmentation of a high energy proton in an intense electromagnetic field produced by the other proton'' seems contradictory. Presence or absence of a soft component (what the Bylinkin exponential responds to) depends on dissociation {\em or not} of projectile protons.

\begin{figure}[h]
	\includegraphics[height=1.65in]{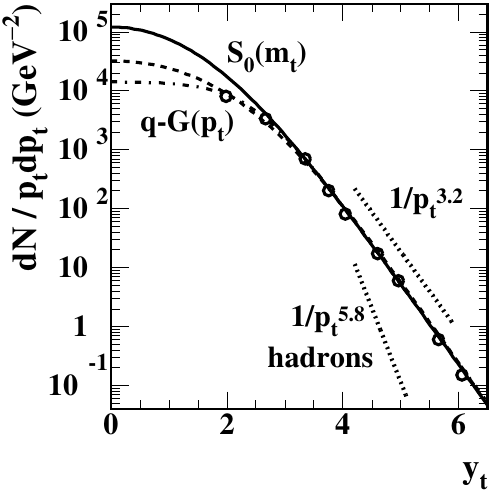}
	\includegraphics[height=1.65in]{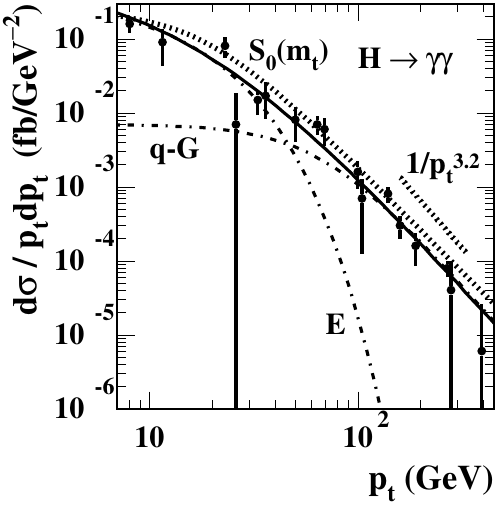}
	\caption{\label{ggmm}
		Left: Dimuon \pt\ spectrum (points) from Fig.~3 of Ref.~\cite{dimaentangle} (points) representing the process $\gamma$-$\gamma \rightarrow \mu^+$-$\mu^-$ for 13 TeV \pp\ collisions. The dash-dotted curve is the Bylinkin power-law model element ($q$-Gaussian) denoted by $q$-G. The solid and dashed curves are TCM model function $\hat S_0(m_t)$ with parameters as described in the text. The lower dotted line represents the power law for hadrons from 13 TeV \pp\ collisions.
		Right: Higgs boson \pt\ spectrum from \mbox{$H\rightarrow \gamma$-$\gamma$} (points). The dash-dotted curves are Bylinkin model elements with parameters adjusted to best accommodate data. The solid curve is TCM model function $\hat S_0(m_t)$ with parameters as described in the text.
}  % acompare9b, 21b
\end{figure}

Figure~\ref{ggmm} (left) shows a dimuon \pt\ spectrum (points) adapted from Fig.~3 of Ref.~\cite{dimaentangle} representing the process \mbox{$\gamma$-$\gamma$} $\rightarrow \mu^+$-$\mu^-$ for 13 TeV \pp\ collisions. Data and curves are plotted in the format of density on \pt\  vs pion transverse rapidity \yt\ with zero lower bound that gives improved access to lowest \pt. The dash-dotted curve is the Bylinkin model with parameters $T_h = 0.45$ GeV and $N = 1.55$ adjusted here to best describe the data. Also shown is TCM model $\hat S_0(m_t)$ (solid) with pion \mt\ and parameters $T = 0.145$ GeV and $n = 3.2$ that should correspond to Bylinkin $N = 1.6$. The dashed curve is $\hat S_0(m_t)$ with parameters $T = 0.245$ GeV and $n = 3.3$. The latter demonstrates that a function with {\em exponential limiting case} is not ruled out by the available data. It should also be noted that projectile protons here {\em remain intact}: there is no dissociation and therefore no parton splitting cascade hadronizing to a TCM soft component.

%%%%%%%%%%
%%%%%%%%%%%%

In contrast to expectations for $\gamma$-$\gamma \rightarrow \mu^+$-$\mu^-$,  spectra for Higgs bosons decaying to $\gamma$-$\gamma$ or to four leptons are expected to {\em include} an exponential component reflecting thermalization if entanglement is involved. Applying the Bylinkin model to a Higgs spectrum in its Fig.~4, Ref.~\cite{dimaentangle} concludes that ``there clearly are both the hard scattering (power law) and thermal (exponential) components in the [\pt\ spectrum] similarly to...'' that for their Fig.~2.

Figure~\ref{ggmm} (right) shows the same Higgs boson spectrum (points) from $H\rightarrow \gamma$-$\gamma$. The dash-dotted curves are elements of the Bylinkin model (as they appear in Ref.~\cite{dimaentangle} Fig.~4) where the exponential element E has $T_{th} = 10$ GeV and the power-law element $q$-G has $T_h = 53$ GeV and $N = 1.6$ (i.e.\ $n = 3.2$). 
The dotted curve is the power-law element alone with $T_h = 15$ GeV and $N = 1.42$ displaced upward for visibility. The solid curve is TCM model  $\hat S_0(m_t)$ with pion mass for \mt\ and with parameters $T = 5$ GeV and $n = 3.2$. These data are not of sufficient quality to distinguish among three models and cannot establish the presence {\em or absence} of a separate exponential function. The same may be said for Fig.~5 of Ref.~\cite{dimaentangle} representing $H\rightarrow 4l$. One may ask what sense does a ``temperature'' value $T_{th} \approx 15$ GeV convey and what does $T_{h} \approx 53$ GeV (``hard scale'') mean?

In response to conjecture in Ref.~\cite{dimalev} that entanglement might lead to thermalization and related predictions -- ``We propose the entanglement entropy as an observable that can be studied in deep inelastic scattering'' --  Ref.~\cite{h1} (H1) responds ``...in order to further understand the role of colour confinement in high energy collisions, it has been suggested that quantum entanglement of partons could be an important probe of the underlying mechanism.... The predictions [derived] from the entropy of gluons are found to grossly disagree with the hadron entropy obtained from the multiplicity measurements presented here, and therefore the data do not support the basic concept of equality of the parton and hadron entropy with the current level of theory development.''

In Ref.~\cite{iskander} quantum entanglement and thermal behavior are further probed with neutrino-induced reactions. The basis for analysis is again the Bylinkin model and the perceived presence or absence of its exponential element in fits to spectra. A contrast is made between (a) interactions where only part of a nucleon is involved (termed differential scattering) and (b) where the entire target nucleon or nucleus is involved (termed coherent scattering). Case (a) is represented by $\bar \nu_\mu + N \rightarrow \mu^+ + \pi^0 + X$ and (b) is represented by $\bar \nu_\mu + A \rightarrow \mu^+ + \pi^- + A$

\begin{figure}[h]
	\includegraphics[height=1.7in]{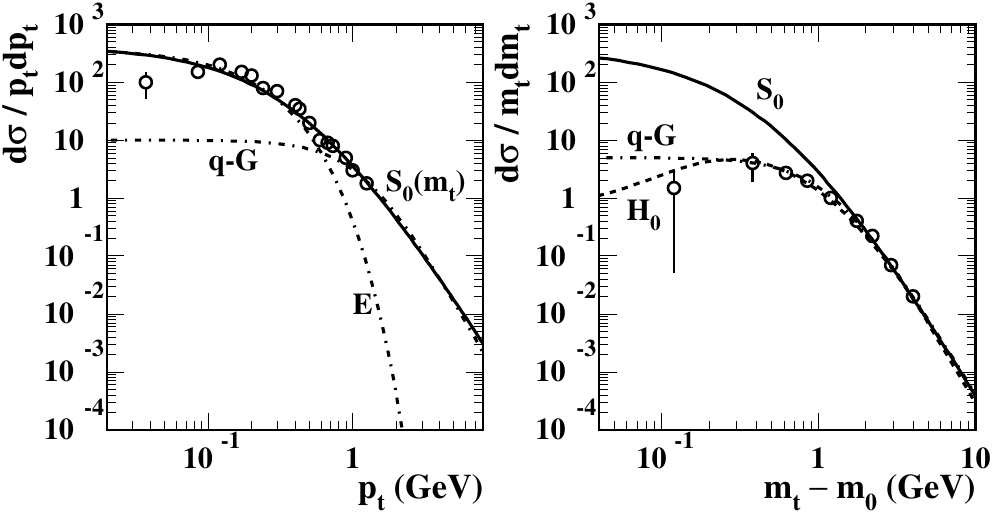}
	\caption{\label{iskander}
		Left:  Data (points) adapted from Fig.~2 of Ref.~\cite{iskander} for the process $\bar \nu_\mu + N \rightarrow \mu^+ + \pi^0 + X$. The dash-dotted curves represent elements of Eq.~(\ref{byeq}). The solid curve is TCM soft-component model $\hat S_0(m_t)$. Parameter values are given in the text.
		Right:  Data (points) adapted from Fig.~3 of Ref.~\cite{iskander} for the process $\bar \nu_\mu + A \rightarrow \mu^+ + \pi^- + A$. The dash-dotted curve is the power-law element of Eq.~(\ref{byeq}). The solid curve is TCM soft component $\hat S_0(m_t)$. The dashed curve is TCM hard component  $\hat H_0(y_t)$ transformed to a density on \mt.
} % acompare22
\end{figure}

Figure~\ref{iskander} (left) shows data (points) adapted from Fig.~2 of Ref.~\cite{iskander} for case (a). The dash-dotted curves represent the elements of Eq.~(\ref{byeq}) with parameters $T_{th} = 0.145$ GeV, $T_h = 0.8$ GeV and $N = 2.2$ adjusted to best reproduce the Ref.~\cite{iskander} Fig.~2 results. The solid curve is $\hat S_0(m_t)$ with parameters $T = 0.145$ GeV and $n$ = 4.6. The two models are indistinguishable and describe the data reasonably well. However, the Bylinkin model requires five fitted parameters whereas $\hat S_0(m_t)$ requires only two, since $T = 0.145$ is predicted for a pion spectrum. There is no possibility from these data to demonstrate that either the $q$-Gaussian or exponential is required. If the Bylinkin exponential were replaced by $\hat S_0(m_t)$, {\em whose limiting case for $1/n \rightarrow 0$ is indeed an exponential}, it is likely that a $\chi^2$ fit would return zero amplitude for the $q$-Gaussian, and Bayesian analysis would reject the composite Bylinkin model  based on needless complexity. There is thus no evidence in these data for a {\em required} exponential element signaling thermalization.

%%%%%%%%%%
%%%%%%%%%%%%

Figure~\ref{iskander} (right) shows data (points) adapted from Fig.~3 of Ref.~\cite{iskander} for case (b).  $q$-Gaussian (dash-dotted) parameters are $T_h = 0.8$ GeV and $N = 2.2$ ($n = 4.4$). TCM soft component $\hat S_0(m_t)$ (solid) parameters are $T = 0.145$ GeV and $n = 4.6$. Hard component $\hat H_0(m_t)$ (dashed) parameters are $\bar y_t =2.65$, $\sigma_{y_t} = 0.9$ and $q = 2.5$ (i.e.\ $n = 4.5$). Recall that $\hat H_0(y_t)$ is a Gaussian on \yt\ with exponential tail transformed here to a density on 
\mt\ via Jacobian $y_t/m_t p_t$. Data prefer the TCM hard component to the Bylinkin $q$-Gaussian. Data appear to reject a soft component which is reasonable since once again the projectile hadrons remain intact in the final state and there is thus no parton splitting cascade.

\subsection{Relation between thermal and hard scales} \label{relation}

Reference~\cite{pajares} addresses the issue of possible thermalization via entanglement within high-energy \pp\ collisions, emphasizing an inferred relation between thermal temperature $T_{th}$ and ``hard scale'' $T_h$. The conjectured thermalization process is described thus: ``In a high-energy collision a hard parton interaction produces a rapid quench of the entangled partonic state [of projectile nucleons] and thus the characteristic effective temperature [$T_{th}$ e.g.\ inferred from the Bylinkin exponential model element] can depend on the energy scale of the hard process... [parameter $T_h$ of the Bylinkin $q$-Gaussian].'' In order to establish the relation between $T_{th}$ and $T_h$ ``we...perform an extensive study of the energy and multiplicity dependence of the hard and soft scales in $pp$ collisions....''  The study asserts that ``...the relation between both scales is determined approximately by the inverse of the normalized fluctuations of the number of partons of the initial wave function or, equivalently, of the normalized fluctuations of the hard scale.'' Results of that study are examined here in  context of the TCM.

\begin{figure}[h]
	\includegraphics[height=1.65in]{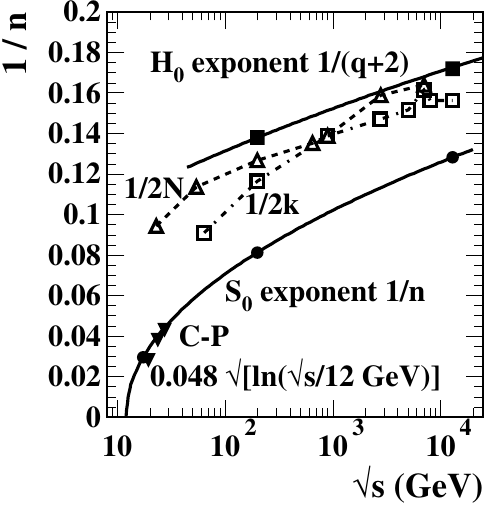}
	\includegraphics[height=1.65in]{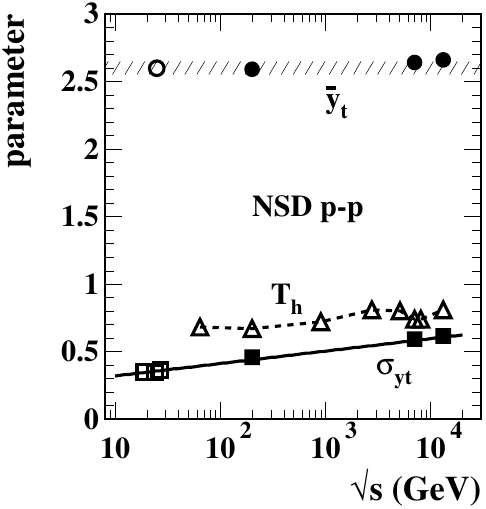}
	\caption{\label{byparams}
		Left: Model exponents plotted as reciprocals. TCM hard component $1/(q+2)$ and TCM soft component $1/n$ from Ref.~\cite{alicetomspec} (solid points) compared with Bylinkin $1/2N$ from Ref.~\cite{bylinkin2010} and equivalent parameter $1/2k$ from Ref.~\cite{pajares} (open points).
		Right: TCM hard-component parameters $\bar y_t$ (mode) and $\sigma_{y_t}$ (width above the mode) from Ref.~\cite{alicetomspec} vs Bylinkin power-law parameter $T_{h}$ ($q$-Gaussian width) from Ref.~\cite{pajares}.
} % acompare11a, 11b
\end{figure}

Figure~\ref{byparams} (left) repeats Fig.~\ref{power} (left) with the addition of values $k$ from Table I of Ref.~\cite{pajares} plotted as $1/2k$ (open squares) to be compatible with the other data. Open triangles represent parameter $N$ taken from  Fig.~3 of Ref.~\cite{bylinkin2015}. Those data are obtained via fits of Eq.~(\ref{byeq}) to charged-hadron spectra from MB or NSD \pp\ collisions. The solid points are derived from the TCM applied to LHC, RHIC, SPS and NAL (fixed-target) \pp\ spectra. Displacement of open points from the upper solid curve (TCM hard component) arises from the Bylinkin $q$-Gaussian performing double duty: not only describing the jet contribution, peaked near 1 GeV/c, but also {\em part of} the nonjet soft component, thus pulling the $1/2N$ or $1/2k$ values down corresponding to a softer tail.

%%%%%%%%%%
%%%%%%%%%%%%

Figure~\ref{byparams} (right), adopted from Ref.~\cite{alicetomspec}, presents energy evolution of TCM hard-component $\hat H_0(y_t)$ parameters indicating that the peak mode $\bar y_t$ is essentially independent of collision energy while the peak width {\em above the mode} $\sigma_{y_t}$ increases linearly according to $\ln(\sqrt{s})$ reflecting evolution of underlying jet spectra. Solid points are from RHIC and LHC while open points at lowest energies are from NAL fixed-target spectra~\cite{tomnmf}. The open triangles are Bylinkin-model power-law parameter $T_{h}$ obtained from Table I of Ref.~\cite{pajares}. The systematic upward displacement of the last arises from the difference in required width $T_h \leftrightarrow \sigma_h$ of a $q$-Gaussian centered at zero \pt\ vs that for a peaked distribution centered near 1 GeV/c. Otherwise, the energy trends agree within uncertainties.

\begin{figure}[h]
	\includegraphics[height=1.65in]{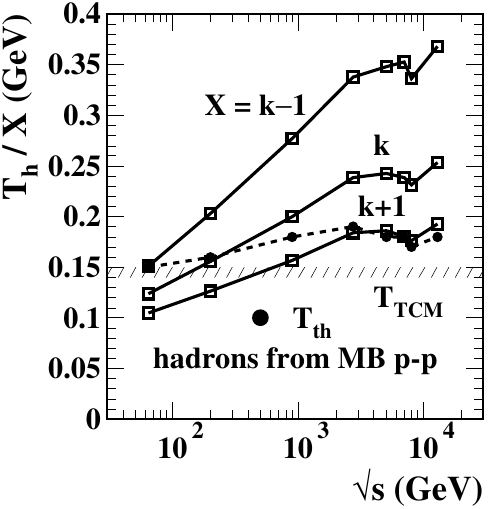}
	\includegraphics[height=1.65in]{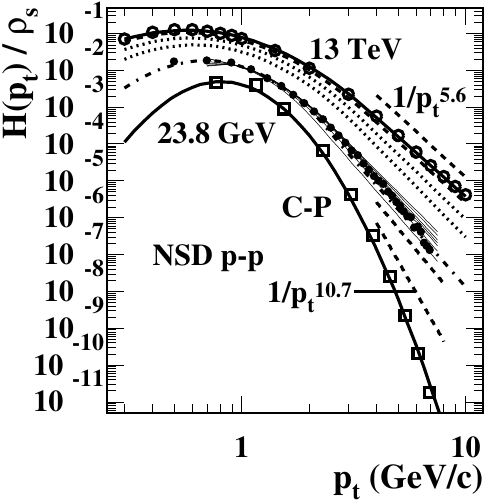}
	\put(-200,103) {$\bf k \leftrightarrow N$}
	\caption{\label{energy}
		Left: Bylinkin-model exponential parameter $T_{th}$ and power-law parameters $T_{pl} / X$ ($X$ representing three instances of power-law exponent $k$) from Ref.~\cite{pajares} vs collision energy. The hatched band represents TCM soft-component $T \approx 0.145$ GeV for unidentified hadrons.
Right: Spectrum hard components over the currently accessible energy range from threshold of dijet production (10 GeV) to LHC top energy (13 TeV). The curves are determined by systematic variation with energy of TCM parameters as in Refs.~\cite{alicetomspec,tomnmf}. The points are from Refs.~\cite{tomnmf} (23.8 GeV),~\cite{ppquad} (200 GeV) and~\cite{alicespec} (13 TeV).
} % acompare11c, alice125dnew
\end{figure}

Figure~\ref{energy} (left) responds to Eq.~(4) of Ref.~\cite{pajares}, a proposed linear relation between ``thermal'' parameter $T_{th}$ and ``hard'' parameter $T_h$ said to arise from ``rapid quench of the entangled partonic state'' as noted above. Open squares represent $T_h$ and $k$ values from Table I of Ref.~\cite{pajares} plotted as $T_h / X$ where $X = k-1,~k$ or $k+1$ to illustrate the consequences of choosing one form over others.  Bylinkin exponential parameter $T_{th}$ increases significantly with energy, but that increase can be attributed to increased jet contribution to spectra at higher \pt\ (see below). Soft-component slope parameter $T_\text{TCM} = 0.145 \pm 0.005$ GeV for unidentified hadrons is represented by the hatched band -- no significant energy dependence. Relations $T_h / X$ on the other hand strongly increase with collision energy; the mechanism should be clear from Fig.~\ref{byparams}: strong increase of jet production. No reason is given for the specific choice of Eq.~(4) in Ref.~\cite{pajares}, but it best accommodates the claimed relation.

%%%%%%%%%%
%%%%%%%%%%%%

Figures~5 and 6 of  Ref.~\cite{pajares} show $T_{th}$ vs $T_h/(k+1)$ comparisons for identified hadrons from \pp\ and \ppb\ collisions vs event multiplicity \nch. The figures suggest confirmation of Eq.~(4) in Ref.~\cite{pajares} given apparent overlap of $T_{th}$ and $T_h/(k+1)$ error bands; however, that can be questioned. TCM analysis of PID spectrum data indicates that while pions consistently require $\hat S_0(m_t)$ $T \approx 0.145$ GeV all heavier hadrons are well-described by $T \approx 0.200$ GeV independent of \nch. See Fig.~\ref{omegarats} (right) for Omegas. That $T_{th}$ might increase substantially with \nch\ and hadron mass could arise because the Bylinkin exponential model element is poorly suited to describe data and must become ``flatter'' with increasing mass and jet production (noted above). And in Refs.~\cite{bylinkinanom,bylinkphen,bylinkintherm} the exponential element and $T_{th}$ are said to be required only for pions, which presents a complication for Ref.~\cite{pajares} Eq.~(4).

The PID trends in  Ref.~\cite{pajares} can be explained as follows. The Bylinkin $T_h$ parameter (width of a $q$-Gaussian) must increase substantially with hadron mass because (a) the shoulder of the data {\em soft} component moves to higher \pt\ with mass as in Fig.~\ref{shape} (right) and (b) the hard/soft fraction ratio $\tilde z_i = z_{hi} / z_{si}$ increases linearly with mass as described in Sec.~\ref{omega} and leads to dominance of jet fragments for higher-mass hadrons as illustrated in Fig.~\ref{omegarats}. The Ref.~\cite{pajares} $k$ parameter ($= N$ for Bylinkin) also increases with hadron mass (the tail softens) again per Fig.~\ref{shape} (right) which is partly offset by the growth of jet production and hard component with \nch. Ratio $T_h/(k+1)$ confuses several effects and is uninterpretable.

Figure~\ref{energy} (right) adopted from Refs.~\cite{alicetomspec,tomnmf} responds to Eq.~(9) of Ref.~\cite{pajares} which relates to the statement: ``...we devise also a convenient and simple parametrization of the whole soft and hard [\pt] spectrum for the full range of energies explored at RHIC and the LHC.'' The procedure starts with  the Bylinkin power-law element, adds fluctuations and obtains a single function with exponential behavior at low \pt\ and power-law at high \pt. The result is shown in its Fig.~8 as applied to several presumably minimum-bias \pt\ spectra where the jet contribution is minimal. The functional form of TCM $\hat S_0(m_t)$ alone as a {\em monolithic} model (i.e.\ no jets) can do as well with much less complexity as demonstrated in Fig.~\ref{ppcompare} (left). 

What such approaches do not do is confront {\em differential evolution} of spectrum shape with event \nch\ or centrality. The result for full spectrum description is demonstrated in Figs.~\ref{none} and \ref{ppcompare} (right) above where the structure of the TCM is {\em required by data}. The TCM then provides the means to isolate accurately the jet contribution as shown in Fig.~\ref{energy} (right), including results from low-energy fixed-target experiments (open squares)~\cite{tomnmf}.

Ironically, while the stated goal in Ref.~\cite{pajares} is to establish a direct connection between soft and hard components, in the end it states that ``...the exact form of the relation between the soft and hard sectors is nevertheless unknown.'' In contrast, the relation between TCM soft and hard components {\em is} established accurately to within data statistical uncertainties. Each spectrum component has its own unique structure, but they are connected by (a) the all-important relation $\bar \rho_h \approx \alpha(\sqrt{s}) \bar \rho_s^2$ that remains valid over the full range of collision energies and by (b) the universality of a parton splitting cascade as viewed perpendicular or parallel to its axis, per Fig.~\ref{eedata}.

%%%%%%%%%%%%%%%%%
\section{Discussion} \label{disc}

Interpretation of the Bylinkin exponential element as an indicator for thermalization within small collision systems is a recurring issue (see Sec.~\ref{small}). ``The `soft' one [exponential element] is ubiquitous in high energy collisions and has the appearance of the thermal spectrum -- but its origin remains mysterious to this day. ...while in nuclear collisions one may expect thermalization to take place, it is hard to believe that thermalization can occur in...DIS or \ee\ annihilation. ... The universal thermal character of hadron [\pt] spectra and abundances in all high energy processes...begs for a theoretical explanation.'' In this section several issues are addressed: (a) the relation of the TCM soft component to dijets arising from \ee\ $\rightarrow$ \qqbar\ collisions, (b) conjectured quantum entanglement as a mechanism for thermalization in small collision systems vs parton splitting cascades in \pp\ inelastic collisions and dijet formation, (c) spectrum model design -- {\em a priori} vs {\em a posteriori} and (d) resonance contributions to pion spectra and so-called ``anomalous behavior'' of thermalization as inferred from the Bylinkin model.

\subsection{Dijet properties and p-p soft component} \label{dipole}

If one accepts that hadron production via fragmentation of energetic leading particles (hadrons or partons) is a dominant process in high-energy nuclear collisions then the process \ee\ $\rightarrow$ \qqbar\ $\rightarrow$ dijet should provide a particularly clear example. The \ee\ data presented below, \pt\ and \pz\ spectra relative to the dijet axis, correspond to an event condition of single dijet production. One should then expect no jet contribution to the \pt\ spectrum (no three-jet events). Within the Bylinkin model context that condition implies only the exponential model element should play a role, no power-law element.

\begin{figure}[h]
	\includegraphics[width=1.65in]{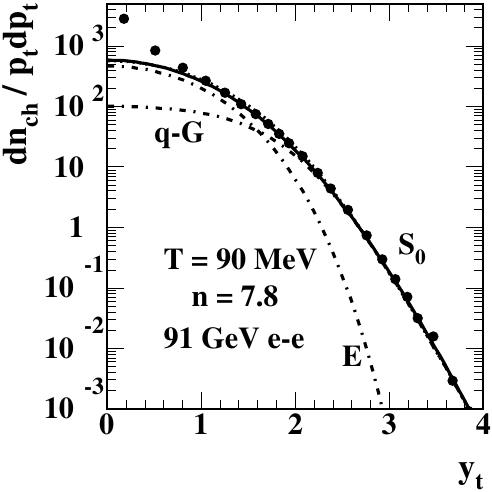}
	\includegraphics[width=1.65in]{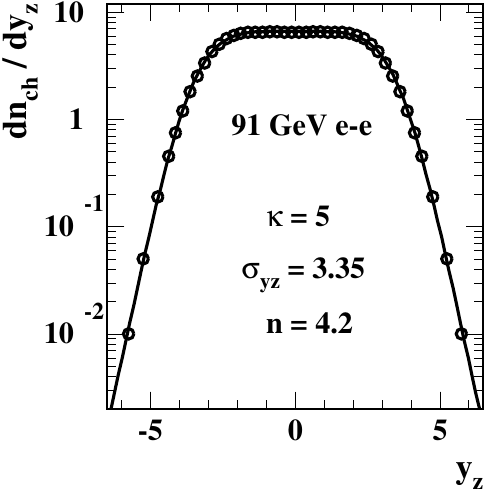}
	\caption{\label{eedata}
		Left: Hadron distribution on \pt\ (perpendicular to the dijet axis) from 91 GeV \ee\ collisions~\cite{alephff} plotted vs pion \yt. The solid curve is TCM model $\hat S_0(m_t)$ with parameters ($T,n$) noted in the figure. The dash-dotted curves are Bylinkin exponential E and $q$-Gaussian \mbox{$q$-G} representing two elements interpreted as thermal and hard contributions. Their sum is dotted and lies on the solid curve.
		Right: Hadron distribution on \yz\ (along the dijet axis) from 91 GeV \ee\ $\rightarrow$ \qqbar\ collisions. The curve is Eq.~(\ref{gcurve}) explained in the text.
	} % aleph2cy,  aleph1new
\end{figure}

Figure~\ref{eedata} (left) shows an \ee\ \pt\ spectrum replotted vs pion \yt\ (derived from measured \pt\ assuming all hadrons are pions as for \yz\ below). The points are published $dn_{ch}/dp_t$ data (Fig.~18a of Ref.~\cite{alephff}) (ALEPH) divided by \pt\ with no other scaling. According to Ref.~\cite{alephff} that is the ``charged particle momentum component transverse to the sphericity [dijet] axis and projected into the event plane.''  The solid curve is soft component $\hat S_0(m_t)$ with slope parameter $T = 90$ MeV and exponent $n = 7.8$. The \ee\ data are described within  uncertainties except for  two points  at 0.02 and 0.07 GeV/c. In comparison, soft-component $\hat S_0(y_t)$ for pions from 13 TeV \pp\ collisions has $T = 145$ MeV and $n = 8.5$~\cite{pppid}.

Ironically, this result deviates dramatically from expectations within the Bylinkin model context: For instance, ``only the inclusive spectra of charged particles produced in pure baryonic collisions require a substantial contribution of the Boltzman[n]-like exponential term~\cite{bylinkin2010}'' and ``...the `thermal' production expressed by the exponential term...is essential only for $pp$- collisions....\cite{bylinkin2014nucphb}.'' But the Bylinkin exponential element is approximating, more or less, the TCM soft-component model in most settings. The dash-dotted curves are the two Bylinkin model elements whose sum approximates the solid curve. The Bylinkin ``thermal'' exponential has $T_{th} = 90$ MeV whereas the ``power-law'' element has $T_h = 0.32$ GeV and $N = 3.3$. The conventional Bylinkin model interpretation implies that the \pt\ spectrum from  \qqbar\ dijets includes a ``hard component'' transverse to the dijet axis.

%%%%%%%%%%
%%%%%%%%%%%%

Figure~\ref{eedata} (right) shows a hadron distribution on \yz\ from \ee\ collisions reported by Ref.~\cite{alephff}. $y_z$  ($y_s$ in Fig.~16a of Ref.~\cite{alephff} with $s$ for ``sphericity'') is derived from measured $p_L \rightarrow p_z$ along the  sphericity (dijet) axis and is based on assigning the pion mass to all hadrons. The solid curve is a {\em generalized} $q$-Gaussian distribution
\bea \label{gcurve}
G(y_z) &=& \frac{A}{[1 + (|y_z|/\sigma_{y_z})^\kappa/n]^{n}}
\\ \nonumber
&\sim & A \exp_q[-(|y_z| /\sigma_{y_z})^\kappa],
\eea
with $\kappa = 5$, $\sigma_{y_z} = 3.35$, $n = 4.2$ and $A = 6.6$.
The  distribution differs from a Gaussian in two ways: (a) the parameter value $\kappa =$ 5 rather than 2 leads to a more rectangular distribution and (b)  parameter value $n =$ 4.2 controls the distribution tails. The $q$ in ``$q$-Gaussian'' refers to relation $q - 1 = 1/n$. For $1/n \rightarrow 0$ (and $\kappa = 2$) the distribution reverts to a standard Gaussian form. 

The ALEPH combination of \pt\ and \pz\ distributions in Fig.~\ref{eedata} provides an alternative to fragmentation functions (FFs) presented as densities on {\em total} momentum $p$ or momentum fraction $x_p = p / p_{jet}$~\cite{eeprd}, the resulting shape sometimes referred to as a ``humpbacked plateau''~\cite{humpback}. Whereas the density on \pz\  is approximately uniform near the origin (\yz\ = 0) the density on {\em total} momentum $p$ falls to zero at the origin, hence the term ``humpbacked.''

Equation~(\ref{gcurve}) can be seen as a generalization of the Bylinkin power-law model element wherein $\kappa \rightarrow 2$ and $n \rightarrow N$. Just as the Bylinkin exponential does not describe the \ee\ \pt\ spectrum, where there is no jet contribution, the Bylinkin power law does not describe  the \pz\ spectrum which is, in a sense, all jet contribution. In \pp\ collisions the \pt\ spectrum soft component is formally equivalent to Fig.~\ref{eedata} (left) described by TCM soft component $\hat S_0(m_t)$ because dissociation of projectile protons along the beam axis is formally equivalent to the parton splitting cascades from \qqbar\ fragmentation. However, the \pt\ spectrum contribution from MB dijets is not equivalent to Fig.~\ref{eedata} (right) since (a) hadron \pt\ relative to the beam axis is not well correlated with individual jet axes and (b) there is a broad underlying jet energy spectrum. Spectrum hard components derived empirically from data as in Fig.~\ref{none} are  shown to be equivalent to a convolution of measured FFs (on total momentum $p$) and a measured jet energy spectrum~\cite{fragevo}. The result is very different from the Bylinkin power-law model element.

\subsection{Entanglement vs parton splitting cascade}

Arguments in Refs.~\cite{dimalev,dimaentangle,bylinkintherm,pajares,iskander} advance a narrative in which apparent thermalization within high-energy collisions, especially in small collision systems, may be explained via quantum entanglement. For example, Ref.~\cite{pajares} asserts that a parton hard scatter may quench an entangled partonic state (within a projectile nucleon), producing a correlation between Bylinkin exponential parameter $T_{th}$ and power-law parameter $T_h$ as noted in Sec.~\ref{relation}. That is the most recent of a series of steps in which (a) certain data features in A-A collisions (e.g.\ jet quenching, elliptic flow) were interpreted to demonstrate quark-gluon plasma (QGP) formation within a thermalized high-density medium, (b) similar data features were observed in small collision systems (LHC 2010 {\em et seq.}) and (c) arguments developed that QGP must also form in small collision systems. The problem then emerges: whence thermalization in small low-density systems?

The Bylinkin spectrum model with its exponential element has emerged as a pivotal actor: The narrative continues that if the exponential model element {\em seems} required by spectra then thermalization has been achieved, presumably via entanglement for small systems. Claims of QGP in small collision systems are then validated. But it is also important to demonstrate where thermalization has {\em not} been achieved, with absence of an exponential element. Then see the several figures in Sec.~\ref{exotic}.

What is missing from that history is an alternative narrative in which (a) certain data features in \aa\ collisions may have been misinterpreted because of incomplete understanding of elementary collisions whether in isolation or as superposed within \aa\ collisions, (b) observation of similar features in small collision systems motivates more-careful study and improved understanding and (c) QCD manifestations in all collision systems are better understood. Equal weight should be given to both narratives until understanding and resolution are achieved.

Interpretation of TCM components has relied on the fundamental element of  parton splitting cascades and universality. The soft component is associated with projectile nucleon dissociation (fragmentation via splitting cascade) along the beam axis which should then be approximately equivalent to a \qqbar\ dijet as in Fig.~\ref{eedata} (left). In neither case is thermalization expected. The hard component is associated with MB dijet production from large-angle scattering of low-$x$ gluons again followed by splitting cascades to form jets. Measured \pp\ spectrum hard components are {\em quantitatively predicted} by convolution of {\em measured} jet energy spectra with {\em measured} fragmentation functions per QCD factorization~\cite{fragevo,eeprd,jetspec2}.

Products of quantum transitions averaged over an event ensemble tend to produce distributions that might be attributed to classical thermalization by extended rescattering. i.e.\ {\em maximum-entropy distributions}. As examples, soft component $\hat S_0(m_t)$ of hadron spectra describing projectile dissociation, the beta distribution that describes parton fragmentation functions~\cite{eeprd} and the Gaussian distribution on jet rapidity that describes jet energy spectra~\cite{jetspec2}. In each case it is highly unlikely that entropy is maximized in the Boltzmann sense to arrive at those distributions. One may instead look to Feynman path integrals and the least action principle: an ensemble of quantum transitions, given fixed constraints, may arrive at maximum-entropy distributions on average. Observation of data structures possibly associated classically with thermalization via multiple scattering does not confirm thermalization in high-energy nuclear collisions.

\subsection{Spectrum model design: {\em a priori} vs data} \label{design}

Spectrum model design requires choices: (a) {\em a priori} argument based on assumptions vs empirical derivation from systematic data trends, (b) number of model elements and parameters, (c) interpretability in terms of comparison with theory and previous experiment.

The choice of specific model elements (functions) is critical for an interpretable model. Consider a space whose points represent possible functions that might be combined to model \pt\ spectra. An exponential function (as in Bylinkin Eq.~(\ref{byeq})) is effectively a point in a subspace spanned by the TCM soft component $\hat S_0(m_t)$ with free parameters $T$ and $n$. As noted, $\hat S_0(m_t)$ becomes exponential $\exp[-(m_t - m_0)/T]$ in the limiting case $1/n \rightarrow 0$. Similarly, the $q$-Gaussian in Eq.~(\ref{byeq}) is a point in a subspace spanned by Eq.~(\ref{gcurve}) with $\kappa \rightarrow 2$ and $\sigma_{y_z} \rightarrow T_h$. As noted above, the corresponding structure in the \pt\ spectrum of \pp\ collisions is described by a peaked distribution with power-law tail and different widths above and below its mode as determined empirically from data. But that is just the TCM hard component $\hat H_0(y_t)$ with its parameters: mode $\bar y_t$, widths $\sigma_{y_t+}$ and $\sigma_{y_t-}$ and exponential tail parameter $q$. If $\bar y_t \rightarrow 0$, $\sigma_{y_t+} \rightarrow T_h$ and $q+2 \rightarrow  N$ the result is approximately Bylinkin $q$-Gaussian on \mt\ (modulo transformation with suitable Jacobian). The Bylinkin model is in effect the result of a choice of {\em specific points in a model space} based on {\em a priori} assumptions.

Particular applications of the Bylinkin model  seem to describe some data adequately but other data may exclude the model. Figures~\ref{eedata} (left), \ref{ppcompare} (left), \ref{ggmm} (right) and \ref{iskander} (left) demonstrate that those data can be adequately represented by $\hat S_0(m_t)$ alone with at most two parameters whereas Eq.~(\ref{byeq}) requires at least four including the amplitude ratio for exponential to $q$-Gaussian components. Those data thus do not {\em require} a Bylinkin-model exponential that might be connected to conjectured thermalization. Figures~\ref{none} and \ref{ppcompare} (right) demonstrate (along with Ref.~\cite{ppprd}) that {\em evolution} of data spectrum shape with event multiplicity {\em requires} a hard component $\hat H_0(y_t)$ and therefore rejects the Bylinkin $q$-Gaussian. Regarding the {\em absence} of an exponential model element, data in Fig.~\ref{ggmm} (left) are not determining whereas Fig.~\ref{iskander} (right) is inconsistent with a ``soft component'' whatever the details. The TCM soft component represents fragments from projectile hadron dissociation or leading-quark fragmentation in \ee\ as in Fig.~\ref{eedata}. If no splitting cascade occurs then there is no soft component. That does not seem relevant to the question of thermalization.

\subsection{Resonance contribution to pion spectra}

In Ref.~\cite{bylinkinanom} it is noted that as a result of PID spectrum fits with the Bylinkin model the exponential (soft) element, attributed to thermalization and prominently present in fits to pion spectra, is much lower {\em or even absent} in fits to heavier hadron spectra, described as ``anomalous behavior.'' As an explanation it is stated that ``It is known that the vast majority of pions in hadronic collisions is produced via multiple cascade decays of the heavier hadronic resonances. This might result in a transformation of some part of the initial power law spectrum of the produced short-lived heavy hadrons to an exponential decay pion distribution.'' Based on TCM analysis it is possible to isolate resonance contributions to pion spectra quantitatively as described below.

\begin{figure}[h]
	\includegraphics[height=1.65in]{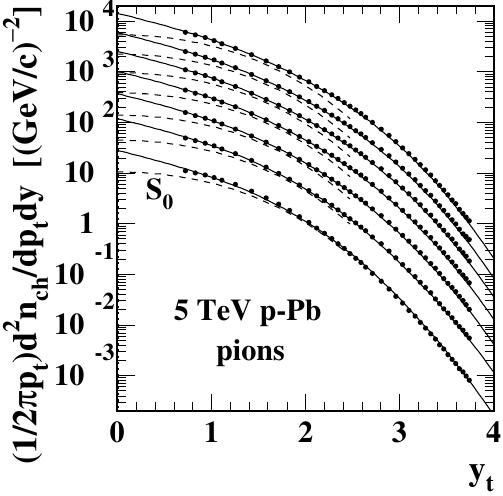}
	\includegraphics[height=1.65in]{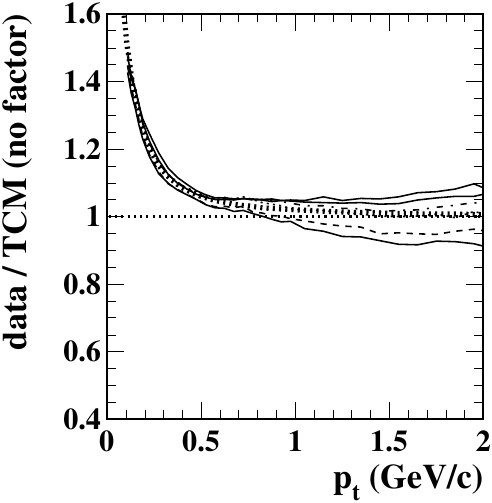}
	\caption{\label{fac1}
		Left: Pion spectrum data (points) and TCM (curves) for seven centralities of 5 TeV \ppb\ collisions. The TCM soft component in this case does not include factor $f(y_t)$ described in the text below.
		Right: Ratios of \ppb\ data to TCM without factor $f(y_t)$ applied to the soft component. The bold dotted curve following data is defined by Eq.~(\ref{bolddotted}).
		}   % alice600aacalx, 600aa22
\end{figure}

Figure~\ref{fac1} (left) shows pion data from 5 TeV \ppb\ collisions (points) and corresponding TCM (curves) including soft-component model $\hat S_{0i}(m_t)$ with predicted model parameters for that collision system~\cite{ppbpid}. Those parameter values are required to describe data spectra above \yt\ = 2 ($p_t \approx 0.5$ GeV/c), but the pion data spectra rise increasingly above the TCM $\hat S_{0i}(m_t)$ below \yt\ = 2.

%%%%%%%%%%
%%%%%%%%%%%%

Figure~\ref{fac1} (right) shows the ratio of spectrum data to TCM as they appear in the left panel for six centrality classes of 5 TeV \ppb\ collisions. Deviations from 1 at higher \pt\ correspond to expected minor evolution of spectrum hard components with \ppb\ centrality. The bold dotted curve is defined by the expression
\bea \label{bolddotted}
f(y_t) &=& 2/\left\{1+\tanh[(y_t - 0.3)/1.2]\right\}.
\eea
For TCM pion spectra to accommodate the resonance contribution to data unit-normal model function $\hat S_0(m_t)$ is replaced by $S_0'(m_t)$ (not normalized) defined by
\bea \label{s0prime}
S_{0i}'(m_t) &=& f(y_t) \hat S_{0i}(m_t).
\eea
In Fig.~\ref{ppcompare} (right) the solid curve at low \pt\ is $S_{0i}'(m_t)$ (resonances) and the dotted curve is $\hat S_{0i}(m_t)$ (none). Note that $\hat S_{0i}(m_t)$ for pions has soft-component ensemble-mean $\bar p_{tsi} \approx 0.44$ GeV/c whereas modified model function $ S_{0i}'(m_t) $ has soft-component mean $\bar p_{tsi} \approx 0.40$ GeV/c due to the additional resonance contribution at lower \pt. Also note $\hat S_0(m_t)$ etc.\ is a density on \mt\ or \pt\ while $f(y_t)$ is a function of pion \yt\ defined in terms of particle \pt.

Quantity $[f(y_t) - 1]\hat S_{0i}(m_t)$ represents a {\em third} component not described by the basic TCM. One may conjecture that this third component is the pion contribution from resonance decays predicted to appear below 0.5 GeV/c~\cite{resonances}. Figure~\ref{fac1} (right) indicates that the resonance yield {\em scales exactly proportional to the soft component} within data uncertainties. Neither the resonance contribution nor TCM soft component $\hat S_{0i}(m_t)$ changes shape significantly with varying \ppb\ centrality. Those results suggest that the resonance contribution has nothing to do with jet production or hard processes.

To summarize, the resonance contribution to pion spectra appears at very low \pt\ with a unique \pt\ dependence that is not exponential (see Eq.~(\ref{bolddotted})). TCM soft component $\hat S_{0i}(m_t)$ is present for all hadron species, with shape evolution depending simply on hadron mass and collision energy. That the Bylinkin exponential might be favored for pion spectra but not more-massive hadrons results from a combination of the resonance contribution at lower \pt\ and movement of the data (and $\hat S_0(m_t)$) soft-component shoulder to higher \pt\ with increasing hadron mass. There is no ``anomalous behavior.''

%%%%%%%%%%%%%%%%
\section{Summary} \label{summ}

This article considers two \pt\ spectrum models applied to data from high-energy \pp\ collisions that are interpreted in quite different ways. A Bylinkin model includes two elements: an exponential function on \mt\ and a ``power-law'' function on \pt\ or \mt\ (actually a Gaussian centered at zero \pt\ with power-law tail). The two elements are interpreted respectively as representing hadron emission from a thermalized source and from a hard-scattering process. In contrast, a two-component (soft + hard) model (TCM)  was inferred empirically from evolution with event multiplicity \nch\ of \pt\ spectra from 200 GeV \pp\ collisions. The two components were interpreted later as arising from projectile-nucleon dissociation (soft) and large-angle scatter of low-$x$ gluons to  jets (hard).

The Bylinkin model has been applied to several collision systems where the apparent presence or absence of the exponential element in fits to data is interpreted to indicate a thermalized particle source or not. Apparent presence of thermalization has subsequently been coupled to conjectures of quantum entanglement as a mechanism for that process. In contrast, the TCM has been applied to many collision systems and several hadron species from which has emerged a coherent picture of collision dynamics generally consistent with conventional QCD theory.

In the present study a detailed comparison is made between the properties of the Bylinkin model and the TCM, for instance trends for parameter variation vs \nch\ and collision energy. Derivation of the TCM by finite differences is presented as an example of lossless data compression wherein the model retains all information carried by particle data. Predictivity of the TCM is demonstrated for the case of high-mass hadron spectra (Cascades, Omegas) with spectra predicted based on the mass dependence of lower-mass spectra (pions, kaons, protons) further illustrated by corresponding ensemble-mean \mmpt\ trends. 

The following  issues emerge from the present study:

(a) The Bylinkin exponential element does not include the ``built-in'' mass dependence of TCM soft component $\hat S_0(m_t)$ demonstrated in this study. As a result, Bylinkin model parameters must vary substantially to accommodate data, but in ways that are not easily interpretable.

(b) Results of Bylinkin model fits to data spectra for varying conditions return parameter trends that strongly disagree with expectations. Examples include soft and hard particle densities vs event \nch\ and collision energies, and corresponding results for ensemble-mean \mmpt.

(c) Given the assumed {\em a priori} constraint on Bylinkin model parameters the two model elements compete to describe the same data.  At lower \pt\ exponential $T_{th}$ and Gaussian width $T_h$ must accommodate soft-component mass dependence and jet/nonjet ratio increasing with event \nch\ and collision energy. At higher \pt\ Gaussian exponent $N$ attempts to accommodate different soft and hard power-law tails again depending on the jet/nonjet ratio. Those issues are camouflaged to an extend if analysis is confined to minimum-bias spectra where the jet contribution is minimal. In contrast, the TCM is required to describe all data from any collision system.

(d) Bylinkin model elements are special cases of more-general functions. The exponential element is  TCM $\hat S_0(m_t)$ with $1/n = 0$ held fixed. The ``power-law'' element is a Gaussian on \pt\ with power-law tail in which the centroid is fixed at zero \pt. A fit to data spectra with both of those parameters {\em freely varying} is likely to return nonzero values for both that closely approximate TCM results by which a broad range of spectrum data are accurately described within statistical uncertainties.
%That subject is addressed in Sec.~\ref{design}.

(e) The energy dependence of TCM $\hat S_0(m_t)$ exponent $n$ that determines its deviation from an exponential is consistent with Gribov diffusion as a fundamental aspect of parton splitting cascades and thus can be extrapolated across a broad range of collision energies and systems. In particular it describes the \pt\ spectrum from \qqbar\ dijets (relative to the dijet axis) within data uncertainties. Exponent $n$ (i.e.\ $1/n$) measures {\em departure} from equilibrium for some systems. It is notable that $n$ is {\em independent of particle density}, again to the limits of data uncertainties.

(f) Given the above issues interpretation of the Bylinkin exponential as signaling thermalization or not depending on model fit results is problematic. TCM soft component $\hat S_0(m_t)$ represents hadrons from longitudinal fragmentation of a leading particle (projectile hadron or quark). If there is no such fragmentation $\hat S_0(m_t)$ is absent. But perceived absence of the Bylinkin exponential in model fits (e.g.\ for more-massive hadron spectra) does not imply absence of projectile-hadron dissociation.

Taken together, the results of the present study suggest that the Bylinkin model is not a suitable model for \pt\ spectra from high-energy nuclear collisions. Physical interpretation of model parameter trends within that model context alone appear quite difficult. In particular the model cannot determine whether a particular data spectrum  reflects thermalization of a particle source.

%%%%%%%%%%%%%%%%

\end{document}